\documentclass[aps,onecolumn,a4paper,superscriptaddress,longbibliography,nofootinbib,notitlepage,groupedaddress]{revtex4-1}

\usepackage{lipsum}
\usepackage{setspace}

\usepackage{graphicx}
\usepackage{amsmath}
\usepackage{amssymb}
\usepackage{amsthm}
\usepackage[utf8]{inputenc}
\usepackage{comment}

\usepackage{color}
\usepackage[dvipsnames]{xcolor}

\usepackage[a4paper,total={6.5in,9in}]{geometry}
\usepackage[english]{babel}
\usepackage[T1]{fontenc}
\usepackage{mathrsfs}
\usepackage{braket}
\usepackage{physics}
\usepackage{enumerate}
\usepackage{graphicx}
\usepackage{mathtools}
\usepackage[export]{adjustbox}
\usepackage{microtype}

\theoremstyle{plain}

\usepackage[colorlinks = true,
            linkcolor = red,
            urlcolor  = blue,
            citecolor = blue,
            anchorcolor = red]{hyperref}

\newtheorem{theorem}{Theorem}
\newtheorem{corollary}[theorem]{Corollary}
\newtheorem{proposition}[theorem]{Proposition}
\newtheorem{lemma}[theorem]{Lemma}

\newtheorem*{remark*}{Remark}   
\renewcommand\qedsymbol{$\blacksquare$}
\newenvironment{proof-of}[1][{\hspace{-\blank}}]{{\medskip\noindent\textit{Proof~{#1}.\ }}}{\hfill\qedsymbol}

\renewcommand{\dim}{{\operatorname{dim}}}

\newcommand{\nc}{\newcommand}
\nc{\rnc}{\renewcommand}

\nc{\avg}[1]{\langle#1\rangle}
\nc{\Rank}{\operatorname{Rank}}
\nc{\smfrac}[2]{\mbox{$\frac{#1}{#2}$}}

\nc{\ox}{\otimes}
\nc{\dg}{\dagger}
\nc{\da}{\downarrow}
\nc{\ua}{\uparrow}
\nc{\cA}{{\cal A}}
\nc{\cB}{{\cal B}}
\nc{\cC}{{\cal C}}
\nc{\cF}{{\cal F}}
\nc{\cG}{{\cal G}}
\nc{\cH}{{\cal H}}
\nc{\cI}{{\cal I}}
\nc{\cJ}{{\cal J}}
\nc{\cK}{{\cal K}}
\nc{\cL}{{\cal L}}
\nc{\cM}{{\cal M}}
\nc{\cN}{{\cal N}}
\nc{\cO}{{\cal O}}
\nc{\cP}{{\cal P}}
\nc{\cQ}{{\cal Q}}
\nc{\cR}{{\cal R}}
\nc{\cS}{{\cal S}}
\nc{\cX}{{\cal X}}
\nc{\cY}{{\cal Y}}
\nc{\cW}{{\cal W}}
\nc{\cZ}{{\cal Z}}
\nc{\SWAP}{{\mathrm{SWAP}}}
\nc{\csupp}{{\operatorname{csupp}}}
\nc{\qsupp}{{\operatorname{qsupp}}}

\begin{document}

\title{R\'enyi-2 entanglement of purification of the symmetric Werner state}

\author{Shokoufe Faraji$^{1,2}$}
\email{s3faraji@uwaterloo.ca}
\affiliation{$^{1}$Department of Physics and Astronomy, University of Waterloo, ON, Canada}
\affiliation{$^{2}$Perimeter Institute for Theoretical Physics, ON, Canada, N2L 2Y5}





\begin{abstract}
This paper determines exactly the Rényi-2 entanglement of purification of the normalized symmetric Werner state in every local dimension $d\ge2$
\begin{equation*}
\mathrm{EoP}_2(W_{+,d})=\log_2\frac{4d(d+1)}{d^2+3d+4}. 
\end{equation*}
We establish additivity for tensor power and, more generally, additivity for finite products of symmetric Werner states with different local dimensions. A canonical purification and its products achieve these values globally, even when arbitrary joint purifications with unrestricted finite dimensional auxiliary systems are allowed. We further examine perturbations away from the symmetric endpoint, and prove the continuity of the optimal value at this endpoint; moreover, we show a sharp square root order of variation in the optimal purity, while the canonical purification remains asymptotically optimal in the achieved entropy as this weight goes to zero.


\end{abstract}
\maketitle

\section{Introduction}

Entanglement of purification measures how much entanglement is needed to represent a mixed state as part of a larger pure state. Given a state on two systems $A$ and $C$, one can consider all possible purifications obtained by adding auxiliary systems, and minimize the entanglement across the enlarged bipartition. The Rényi-2 version replaces the usual entropy by

\begin{equation*}
S_2(\rho)=-\log_2\operatorname{Tr}(\rho^2).
\end{equation*}
This makes the problem closely connected to output purity, since minimizing $S_2$ is equivalent to maximizing $\operatorname{Tr}(\rho^2)$.
Entanglement of purification was introduced by Terhal, Horodecki, Leung, and DiVincenzo \cite{10.1063/1.1498001}, who also studied two qubit Werner states numerically. The Werner states themselves form a wellknown family of highly symmetric bipartite states \cite{PhysRevA.40.4277}.  For the von Neumann entanglement of purification, Christandl and Winter obtained exact results for states supported on symmetric or antisymmetric subspaces \cite{1499048}, also Chen and Winter later presented numerical evidence for nonadditivity in two qubit Werner states \cite{chen2012nonadditivityentanglementpurificationbeyond}. More recently, nonadditivity has been established for the von Neumann entanglement of purification of certain two-qubit Werner states \cite{BaghaliKhanianNegari2026}, while related results for Rényi entanglement of purification have shown that additivity can depend on the Rényi order and the state family \cite{NegariBaghaliKhanian2026}.

This work most directly related to approach of \cite{faraji2026additivityresultsrenyi2entanglement}, whom introduced the constrained Schatten norm formulation of Rényi entanglement of purification and proved multiplicativity results for transpose depolarizing channels and their complementary channels. This also identified the adjoint of the complementary channel associated with Werner states and pointed out that its multiplicativity requires a separate analysis. In the present work, we resolve this problem at the symmetric endpoint $(t=1/(d+1))$. Our argument establishes an exact norm bound for the corresponding map and its tensor powers, without restricting the joint postprocessing channels.


The main difficulty is that the optimization is over all finite dimensional purifications. The auxiliary systems are not fixed in advance, and for several copies the purification may correlate all copies jointly. Therefore, one cannot assume that an optimal purification has the same symmetry as the original state, is a product across copies, or comes from a specially chosen class of channels.
We study this problem for the normalized symmetric Werner state ${W_{+,d}}$. This family is simple enough that its symmetry can be used explicitly, but still the purification optimization remains nontrivial because the auxiliary degrees of freedom are unrestricted.

The strategy is to replace the purification problem by an equivalent optimization over quantum channels acting on a reference system. That channel optimization is then reduced to a concrete operator inequality. The proof is constructive: we first derive the constraints satisfied by every admissible channel, then build an upper bound that applies to all of them, and finally show a channel that reaches that bound. The problem is to determine \begin{equation}
\mathrm{EoP}_2(A:C)_{W_{+,d}},
\end{equation}
and to decide whether the value is additive under tensor powers
\begin{equation}
\mathrm{EoP}_2\left(A^{\otimes n}:C^{\otimes n}
\right)_ {W_{+,d}^{\otimes n}}\stackrel{?}{=}
n\, \mathrm{EoP}_2(A:C)_{W_{+,d}}.
\end{equation}
The optimization on the left allows an arbitrary finite dimension purification of $W_{+,d}^{\otimes n}$, including purifications that correlate all $n$ copies.
We prove the exact one copy value

\begin{equation}
\mathrm{EoP}_2(A:C)_{W_{+,d}}= \log_2\left(\frac{4d(d+1)}{d^2+3d+4}\right)
\end{equation}
and, for every integer $n\ge1$,

\begin{equation}
\mathrm{EoP}_2\left(A^{\otimes n}:C^{\otimes n}\right)_
{W_{+,d}^{\otimes n}}=n\,\mathrm{EoP}_2(A:C)_{W_{+,d}}.
\end{equation}
The proof also gives an explicit purification that attains the minimum. For $n$ copies, the product of these one copy purifications remains optimal even when compared with arbitrary joint purifications. Therefore, product processing is sufficient to attain the optimum, but it is not assumed in deriving the upper bound.

The argument does not require a separate compactness assumption for unrestricted auxiliary systems. The upper bound is proved uniformly for every finite dimension auxiliary output, while the explicit construction reaches that bound in a finite dimension. Likewise, no covariance assumption, purity assumption on the associated channel, or product assumption on the optimizer is used. While the construction identifies an optimal purification, it does not classify all optimal purifications, uniqueness of the optimizer and minimality of the displayed auxiliary dimensions, they are separate questions and are not needed for either the exact value or the additivity result. Finally, the same method also extends to finite products of normalized symmetric Werner states with different local dimensions. What is proved here is therefore an exact result for this symmetric Werner family. 

The paper is organized as follows. Section \ref{sec:2:sym-werner} introduces the symmetric Werner state and its associated map. Section \ref{sec:3:reduction} reduces the optimization to the symmetric input space. Sections \ref{sec:4:lower-bound} and \ref{sec:5:upper-bound}  establish the matching lower and upper bounds for the one copy problem, respectively. Section \ref{sec:6:tensor-multipl} proves multiplicativity under arbitrary tensor powers. \ref{sec:7:eop.additiv} derives the exact Rényi-2 entanglement of purification and its additivity, including products with different local dimensions. Section \ref{sec:8:stability} studies stability under perturbations away from the symmetric endpoint. Finall, Section \ref{sec:9:sum-con} summarizes the results and conclusions.




\section{The symmetric Werner state and its associated map}\label{sec:2:sym-werner}

We fix an integer $d\ge2$ and $d$ dimensional systems

\begin{equation}
A\simeq C\simeq B_1\simeq B_2\simeq\mathbb C^d, \qquad
H:=B_1\otimes B_2.
\end{equation}
We use fixed orthonormal bases to identify these spaces and all transposes are taken in these bases.  For two $d$ dimensional systems $X,Y$, let $\Pi_{XY}$ be the swap operator,
\begin{equation}
\Pi_{XY}\bigl(|i\rangle_X\otimes|j\rangle_Y\bigr)
=|j\rangle_X\otimes|i\rangle_Y,
\end{equation}
and define the projector onto the symmetric subspace by
\begin{equation}\label{eq:sym-state}
P_+^{XY}:=\frac{I_{XY}+\Pi_{XY}}2.
\end{equation}
Now on $H=B_1\otimes B_2$, we write
\begin{equation}
\Pi:=\Pi_{B_1B_2}, \qquad P_+:=P_+^{B_1B_2},
\qquad \mathsf H_+:=\operatorname{ran}P_+.
\end{equation}
Therefore, $\mathsf H_+\simeq\operatorname{Sym}^2(\mathbb C^d)$ has dimension
\begin{equation}
D_d:=\dim\mathsf H_+=\frac{d(d+1)}2.
\end{equation}
The state considered throughout this section is the normalized projector onto the symmetric subspace of $A\otimes C$:
\begin{equation}
W_{+,d}:=\frac{P_+^{AC}}{D_d}=\frac{I_{AC}+\Pi_{AC}}{d(d+1)}.
\end{equation}
It is the endpoint $t=t_+$ of the Werner family
\begin{equation}
W_d(t)=\frac{1-t}{d^2}I_{AC}+ \frac{t}{d}\Pi_{AC},
\qquad t_+:=\frac1{d+1},
\end{equation}
so that $W_{+,d}=W_d(t_+).$ The reason for introducing this parametrization is that $W_d(t)$ is the normalized Choi state of the transposed depolarizing channel $\Gamma_t: \mathcal L(A)\longrightarrow\mathcal L(C),$ with
\begin{equation}
\Gamma_t(X)=tX^T+(1-t)\operatorname{Tr}(X)\frac{I_C}{d}.
\end{equation}
Indeed, if we take
\begin{equation}
|\Phi_d\rangle=\frac1{\sqrt d}
\sum_{i=1}^d |i\rangle_A\otimes|i\rangle_{A'},
\end{equation}
and $\Phi_d:=|\Phi_d\rangle\langle\Phi_d|$, then, after identifying $A'\simeq C$, we obtain

\begin{equation}
(\operatorname{id}_A\otimes\Gamma_t)(\Phi_d)=W_d(t).
\end{equation}
At the endpoint $t=t_+$, a convenient complementary channel representation is known from the work of Datta-Fukuda-Holevo and Holevo
\cite{DattaFukudaHolevo2006,Holevo2007ComplementaryChannels}. It gives $\Gamma_{t_+}^c:
\mathcal L(A)\longrightarrow\mathcal L(H),$ with

\begin{equation}
\Gamma_{t_+}^c(X)=\frac{2}{d+1}P_+(X\otimes I_{B_2}) P_+,
\end{equation}
where $X$ acts on $B_1$ through the fixed identification $A\simeq B_1$.

Now taking the Hilbert-Schmidt adjoint reverses the direction of the map:
\begin{equation}
(\Gamma_{t_+}^c)^\dagger:
\mathcal L(H)\longrightarrow\mathcal L(A).
\end{equation}
And from the trace pairing identity for the partial trace,
\begin{equation}
(\Gamma_{t_+}^c)^\dagger(Y)=
\frac{2}{d+1}\operatorname{Tr}_{B_2}(P_+YP_+).
\end{equation}
We now scale this adjoint by $d$ and define

\begin{align}\label{eq:main-map}
&\mathcal M_{+,d}:=d\,(\Gamma_{t_+}^c)^\dagger:
\mathcal L(H)\longrightarrow\mathcal L(A),\\
&\mathcal M_{+,d}(Y)=\frac{2d}{d+1}
\operatorname{Tr}_{B_2}(P_+YP_+).
\end{align}
The factor $d$ is chosen for the normalization used in the constrained optimization below. Since

\begin{equation}
\operatorname{Tr}_{B_2}P_+=\frac{d+1}{2}I_{B_1},
\end{equation}
we obtain
\begin{equation}
\begin{aligned}
\mathcal M_{+,d}\left(\frac{I_H}{d^2}
\right)&=\frac{2d}{d+1}\operatorname{Tr}_{B_2}\left(P_+\frac{I_H}{d^2}P_+\right)\\
&=\frac{2d}{d+1}\frac1{d^2}\operatorname{Tr}_{B_2}P_+\\
&=\frac{I_A}{d}.
\end{aligned}
\end{equation}
Then $\mathcal M_{+,d}$ sends the maximally mixed input on $H$ to the maximally mixed state on $A$. Note that the map is completely positive, but it is not trace preserving on arbitrary inputs, since
\begin{equation}
\operatorname{Tr}\mathcal M_{+,d}(Y)
=\frac{2d}{d+1}\operatorname{Tr}(P_+Y).
\end{equation}
The distinction is important: $\mathcal M_{+,d}$ is not a quantum channel on all of $\mathcal L(H)$, but interstingly it has exactly the normalization required on the fixed marginal that will appear in the constrained norm. This suggests restricting the input states to those whose marginal on $H$ is fixed to $I_H/d^2$. Therefore, on this restricted set, applying $\mathcal M_{+,d}$ produces normalized output states. The next subsection defines that constrained optimization.

\subsection{The constrained norm and its tensor powers}

Let $\Omega:\mathcal L(X)\longrightarrow\mathcal L(Y)$ be completely positive and need not be trace preserving. Let $E$ be any finite dimension auxiliary system, and consider positive operators $\sigma_{XE}\in\mathcal L(X\otimes E)$ with fixed marginal set by

\begin{equation}\label{eq:extension-set_F}
\mathcal F_X(E):=\left\{\sigma_{XE}\ge0:
\operatorname{Tr}_E\sigma_{XE}=\frac{I_X}{m}\right\}.
\end{equation}
where $m:=\dim X.$ Every $\sigma_{XE}\in\mathcal F_X(E)$ is automatically a state \cite{faraji2026additivityresultsrenyi2entanglement}. We then define the constrained output norm

\begin{equation}
\upsilon_2(\Omega):=\sup_{\substack{E\ \mathrm{finite\text{ }dimensional}\\
\sigma_{XE}\in\mathcal F_X(E)}}\left\|
(\Omega\otimes\operatorname{id}_E)(\sigma_{XE})\right\|_2.
\end{equation}
Thus the optimization allows arbitrary correlations between $X$ and an auxiliary system $E$, while the marginal on $X$ is fixed. For the map of interest \eqref {eq:main-map}, we have $\dim H=d^2$. By using the normalization of $\mathcal M_{+,d}$,

\begin{align}\operatorname{Tr}_E
\left[(\mathcal M_{+,d}\otimes\operatorname{id}_E)(\sigma_{HE})\right]&=\mathcal M_{+,d}\left(\operatorname{Tr}_E\sigma_{HE}\right)\nonumber\\
&=\frac{I_A}{d}.
\end{align}
Therefore every admissible output is a normalized state. In particular,
\begin{equation}
\left\|(\mathcal M_{+,d}\otimes\operatorname{id}_E)(\sigma_{HE})\right\|_2^2
\end{equation}
is its purity.

Now let $X'\simeq X$ be a reference copy of $X$, with the corresponding fixed basis, and define the normalized maximally entangled state

\begin{equation}
|\Phi_m\rangle_{XX'}:=\frac1{\sqrt m}\sum_{i=1}^m
|i\rangle_X\otimes|i\rangle_{X'},\qquad
\Phi_m^{XX'}:=|\Phi_m\rangle\langle\Phi_m|.
\end{equation}
Now let $\Lambda: \mathcal L(X') \longrightarrow\mathcal L(E)$ be a completely positive trace preserving map. Applying $\Lambda$ to the reference half gives
\begin{equation}\label{eq:sigma_XE-Lambda}
\sigma_{XE}=(\operatorname{id}_X\otimes\Lambda)(\Phi_m^{XX'}),
\end{equation}
with marginal on $X$ as $I_X/m$. Conversely, as every state satisfying \eqref{eq:extension-set_F} arises in this way from a unique CPTP map $\Lambda$. This is the standard Choi correspondence \cite{Choi1975}. Thus, for each fixed $E$, we have
\begin{equation}
\sigma_{XE}\in\mathcal F_X(E)\quad\Longleftrightarrow\quad
\sigma_{XE}=(\operatorname{id}_X\otimes\Lambda)(\Phi_m^{XX'})
\end{equation}
This changes only the parametrization of the optimization; it does not restrict the allowed inputs. We also define the normalized Choi operator of $\Omega$,

\begin{equation}\label{eq:Theta_Omega}
\Theta_\Omega:=(\Omega\otimes\operatorname{id}_{X'})(\Phi_m^{XX'}).
\end{equation}
By Equations \eqref{eq:sigma_XE-Lambda} and \eqref{eq:Theta_Omega} the corresponding output becomes

\begin{equation}
\begin{aligned}
(\Omega\otimes\operatorname{id}_E)(\sigma_{XE})
&=(\Omega\otimes\Lambda)(\Phi_m^{XX'})\\
&=(\operatorname{id}_Y\otimes\Lambda)\left[
(\Omega\otimes\operatorname{id}_{X'})(\Phi_m^{XX'})\right]\\
&=(\operatorname{id}_Y\otimes\Lambda)(\Theta_\Omega).
\end{aligned}
\end{equation}
So finally the constrained norm can equivalently be written as
\begin{equation}
\upsilon_2(\Omega)=\sup_{\substack{
E\ \mathrm{finite\text{ }dimensional}\\
\Lambda:\mathcal L(X')\to\mathcal L(E)\ \mathrm{CPTP}
}}\left\|(\operatorname{id}_Y\otimes\Lambda)
(\Theta_\Omega)\right\|_2.
\end{equation}
This is the form we will use later: the operator $\Theta_\Omega$ is fixed, while all freedom in the optimization is contained in the CPTP map $\Lambda$.

Note that the complementary maps preserve the constrained norm \cite[Proposition~35]{faraji2026additivityresultsrenyi2entanglement},
\begin{equation}
\upsilon_2(\Gamma_{t_+})=\upsilon_2(\Gamma_{t_+}^c). \end{equation}
This does not justify replacing $\Gamma_{t_+}^c$ by its Hilbert-Schmidt adjoint.
In general, $\Omega\longmapsto\Omega^\dagger$
does not preserve $\upsilon_2$. Thus the definition
 \eqref{eq:main-map} should not be viewed as a norm preserving passage from
$\Gamma_{t_+}$ or $\Gamma_{t_+}^c$. The factor $d$ arises instead from the normalized Choi representation of the reduced purification state. By homogeneity, we have 
\begin{equation}
\upsilon_2(\mathcal M_{+,d})
= d\, \upsilon_2 \left( (\Gamma_{t_+}^c)^\dagger \right). \end{equation}


For tensor powers we use the induced product bases
and group all input factors, and therefore
$\Phi_{m^n}=\Phi_m^{\otimes n}$ and
$\Theta_{\Omega^{\otimes n}}=\Theta_\Omega^{\otimes n}$. The optimization becomes

\begin{equation}
\upsilon_2(\Omega^{\otimes n})=\sup_{\substack{
E\ \mathrm{finite\text{ }dimensional}\\
\Lambda^{(n)}:\mathcal L((X')^{\otimes n})\to
\mathcal L(E)\ \mathrm{CPTP}}}
\left\|(\operatorname{id}_{Y^{\otimes n}}\otimes
\Lambda^{(n)})(\Theta_\Omega^{\otimes n})\right\|_2.
\end{equation}
Note that the superscript $(n)$ labels a channel acting on the full $n$ copy reference system, and it does not mean that
\begin{equation}
\Lambda^{(n)}=\Lambda_1\otimes\cdots\otimes\Lambda_n.
\end{equation}
For $\mathcal M_{+,d}$, the reference input is $H'=B_1'\otimes B_2',$ so the allowed joint post-processing maps are

\begin{equation}
\Lambda^{(n)}:\mathcal L((H')^{\otimes n})
\longrightarrow \mathcal L(E).
\end{equation}
The question is therefore whether allowing such arbitrary joint processing can improve on the product of the one copy optimization. Equivalently
\begin{equation}
\upsilon_2(\mathcal M_{+,d}^{\otimes n})=
\bigl(\upsilon_2(\mathcal M_{+,d})\bigr)^n\qquad(n\ge1).
\end{equation}
Of course the lower bound is immediate: taking a product of one copy admissible inputs gives
\begin{equation}
\upsilon_2(\mathcal M_{+,d}^{\otimes n})\ge\bigl(
\upsilon_2(\mathcal M_{+,d})\bigr)^n.
\end{equation}
The real problem is the opposite inequality, because the optimization also allows correlated inputs. The rest of the argument is devoted to proving that this extra freedom does not increase the optimum.


\section{Reduction to the symmetric input space}\label{sec:3:reduction}

The formula for $\mathcal M_{+,d}$ contains the symmetric projector on both sides of its input \eqref{eq:main-map}. It is therefore natural to replace the full input space $H$ by the smaller space $\mathsf H_+$. We must, however, keep track of this change of dimension because the constrained norm fixes the input marginal to the maximally mixed state of the relevant input space. Now let
\begin{equation}
J_+:\mathsf H_+\hookrightarrow H
\label{eq:sym-inclusion}
\end{equation}
be the canonical inclusion. Its adjoint $J_+^\dagger:H\to\mathsf H_+$ projects a vector in $H$ onto the symmetric subspace, so
\begin{equation}
J_+^\dagger J_+=I_{\mathsf H_+},\qquad
J_+J_+^\dagger=P_+.
\label{eq:inclusion-id}
\end{equation}
For an operator $Y\in\mathcal L(H)$,
\begin{equation}
J_+^\dagger YJ_+ \in \mathcal L(\mathsf H_+)
\end{equation}
is therefore the compression of $Y$ to the symmetric subspace. We now define the channel that acts directly on this smaller input space

\begin{equation}\label{eq:sym-partial-tr}
\mathcal R_{+,d}: \mathcal L(\mathsf H_+)
\longrightarrow \mathcal L(A), \qquad
\mathcal R_{+,d}(X) := \operatorname{Tr}_{B_2}
(J_+XJ_+^\dagger),
\end{equation}
where $B_1$ is identified with $A$. The map first embeds $X$ into $B_1\otimes B_2$ and then discards $B_2$. Therefore, $\mathcal R_{+,d}$ is completely positive and trace preserving; this is the standard construction of a quantum channel from an isometry followed by a partial trace \cite{Watrous2018TheTheoryofQuantumInformation,Wilde2017QuantumInformationTheory}. The relation between $\mathcal M_{+,d}$ and $\mathcal R_{+,d}$ is now explicit. By using \eqref{eq:main-map} and relations \eqref{eq:inclusion-id}, therefore

\begin{equation}
\mathcal M_{+,d}(Y)=\frac{d^2}{D_d}\,
\mathcal R_{+,d}(J_+^\dagger YJ_+).
\label{eq:map-compr}
\end{equation}
where since $D_d=d(d+1)/2,$ we have $d^2/D_d=2d/(d+1)$; below it will cancel exactly against the corresponding normalization change in the maximally entangled state.

\medskip

Now to compare the two constrained norm optimizations, we next compare the normalized Choi operators of $\mathcal M_{+,d}$ and $\mathcal R_{+,d}$. Let choose the following real orthonormal basis of $\mathsf H_+$

\begin{equation}
\{|e_a\rangle\}_{a=1}^{D_d}=\{|i,i\rangle:1\le i\le d\}\cup\left\{\frac{|i,j\rangle+|j,i\rangle}{\sqrt2}:1\le i<j\le d\right\}.
\label{eq:sym-basis}
\end{equation}
and let $\mathsf H_+'\subset H'$ be the corresponding copy of this symmetric subspace in the reference system, with basis $\{|e_a\rangle_{\mathsf H_+'}\}_{a=1}^{D_d},$ and define its canonical inclusion by
\begin{equation}
J_+':\mathsf H_+'\hookrightarrow H'
\label{eq:ref-incl}
\end{equation}
In fact, when the first half of the maximally entangled vector is projected onto $\mathsf H_+$, the same basis vectors appear on the reference side without complex conjugation, then the choice of a real basis is useful at exactly one point. Indeed,

\begin{equation}
|\Phi_{d^2}\rangle_{HH'}=\frac1d
\sum_{\mu=1}^{d^2}|\mu\rangle_H\otimes|\mu\rangle_{H'}.
\end{equation}
Applying $J_+^\dagger$ to the first factor gives
\begin{align}
(J_+^\dagger\otimes I_{H'}) |\Phi_{d^2}\rangle &=\frac1d
\sum_{\mu=1}^{d^2} J_+^\dagger|\mu\rangle_H \otimes
|\mu\rangle_{H'}\nonumber
\\
&= \frac1d \sum_{a=1}^{D_d}|e_a\rangle_{\mathsf H_+}
\otimes J_+'|e_a\rangle_{\mathsf H_+'}.
\end{align}
On the other hand,
\begin{equation}\label{eq:Phi-in-basis}
|\Phi_{D_d}\rangle_{\mathsf H_+\mathsf H_+'}
= \frac1{\sqrt{D_d}} \sum_{a=1}^{D_d} |e_a\rangle_{\mathsf H_+} \otimes |e_a\rangle_{\mathsf H_+'}.
\end{equation}
Therefore
\begin{equation}
(J_+^\dagger\otimes I_{H'})|\Phi_{d^2}\rangle_{HH'}
=\sqrt{\frac{D_d}{d^2}}\,(I_{\mathsf H_+}\otimes J_+')
|\Phi_{D_d}\rangle_{\mathsf H_+\mathsf H_+'}.
\label{eq:project-ent-v}
\end{equation}
This is where the change of dimension appears in the Choi state. Finally, by considering \eqref{eq:Theta_Omega} and \eqref{eq:map-compr}, we obtain

\begin{align}
\Theta_{\mathcal M_{+,d}}
&=(\mathcal M_{+,d}\otimes\operatorname{id}_{H'})(\Phi_{d^2})\nonumber\\
&=\frac{d^2}{D_d}(\mathcal R_{+,d}\otimes\operatorname{id}_{H'})\left[
(J_+^\dagger\otimes I_{H'})\Phi_{d^2}(J_+\otimes I_{H'})\right].
\end{align}
Taking the outer product of Equation \eqref{eq:project-ent-v} contributes the factor $\frac{D_d}{d^2}.$ It therefore cancels the prefactor $d^2/D_d$ above, and we obtain
\begin{equation}
\Theta_{\mathcal M_{+,d}}=(I_A\otimes J_+')
\Theta_{\mathcal R_{+,d}}(I_A\otimes J_+'^\dagger).
\label{eq:supp_reduc}
\end{equation}
Thus the Choi operator of $\mathcal M_{+,d}$ contains no additional part outside the embedded reference space $\mathsf H_+'$. The change from $H$ to $\mathsf H_+$ changes the normalization of the map and of the maximally entangled vector in opposite ways, and these factors cancel exactly.

It only remains to check that restricting the reference system from $H'$ to $\mathsf H_+'$ does not remove any attainable post-processed output. For $n\ge1$, define the inclusion of the $n$ copy symmetric reference space by

\begin{equation}
J_{+,n}' := (J_+')^{\otimes n}, \qquad
P_{+,n}' := J_{+,n}'J_{+,n}'^\dagger.
\label{eq:tensor-ref-inclu}
\end{equation}
where 
\begin{equation}
J_{+,n}': (\mathsf H_+')^{\otimes n} \hookrightarrow (H')^{\otimes n},
\end{equation}
and $P_{+,n}'$ is the projector onto its image. First let take an arbitrary joint CPTP map

\begin{equation}
\Lambda^{(n)}:\mathcal L((H')^{\otimes n}) \longrightarrow
\mathcal L(E).
\end{equation}
Because the Choi operator in Equation \eqref{eq:supp_reduc} is supported only on the embedded symmetric reference space, therefore only the action of $\Lambda^{(n)}$ on that subspace can affect the output. Therefor, it is natural to define its restriction by

\begin{equation}
\Lambda_+^{(n)}(Z):=\Lambda^{(n)}
(J_{+,n}'ZJ_{+,n}'^\dagger),\qquad
Z\in \mathcal L((\mathsf H_+')^{\otimes n}).
\label{eq:restricted-postpro}
\end{equation}
This map is CPTP because it is a composition of completely positive maps, also trace preserving

\begin{equation}
\begin{aligned}
\operatorname{Tr}\Lambda_+^{(n)}(Z)
&=\operatorname{Tr}(ZJ_{+,n}'^\dagger J_{+,n}')=
\operatorname{Tr}Z.
\end{aligned}
\end{equation}
Tensor version of Equation \eqref{eq:supp_reduc} gives

\begin{equation}
\begin{aligned}\label{eq:matched-postp-out}
(\operatorname{id}_{A^{\otimes n}}
\otimes\Lambda^{(n)})
\left( \Theta_{\mathcal M_{+,d}}^{\otimes n}
\right) = (\operatorname{id}_{A^{\otimes n}}
\otimes\Lambda_+^{(n)})
\left(\Theta_{\mathcal R_{+,d}}^{\otimes n}\right).
\end{aligned}
\end{equation}
So every output attainable in the $\mathcal M_{+,d}^{\otimes n}$ optimization is also attainable in the reduced space optimization. We also need the converse, then let suppose

\begin{equation}
\Lambda_+^{(n)}:\mathcal L((\mathsf H_+')^{\otimes n})
\longrightarrow \mathcal L(E)
\end{equation}
is an arbitrary CPTP map. We want a CPTP map on the full reference space whose action agrees with $\Lambda_+^{(n)}$ on the symmetric subspace. We start by choosing a unit vector $|e_0\rangle\in E$ and define

\begin{equation}
\begin{aligned}
\widehat\Lambda^{(n)}(Z)&:=\Lambda_+^{(n)}
(J_{+,n}'^\dagger ZJ_{+,n}') +\operatorname{Tr}
\left[\left(I_{(H')^{\otimes n}}-P_{+,n}'\right)Z\right]
|e_0\rangle\langle e_0|.
\end{aligned}
\label{eq:extend-postpr}
\end{equation}
The first term applies $\Lambda_+^{(n)}$ to the part of $Z$ lying in the symmetric reference space. The second term handles the orthogonal complement naturally, since simply it throws away all information in the orthogonal complement and replaces it by one fixed state $|e_0\rangle\langle e_0|$ and only the total weight of $Z$ outside the symmetric subspace matters. Therefore, this term restores precisely the trace that was lost when we kept only the symmetric part. Such measure and prepare maps are completely positive; this is a standard channel construction \cite{Watrous2018TheTheoryofQuantumInformation,Wilde2017QuantumInformationTheory}. Now the two terms together preserve trace. Indeed,

\begin{align}
\operatorname{Tr}\widehat\Lambda^{(n)}(Z)
&=\operatorname{Tr}(P_{+,n}'Z)+\operatorname{Tr}
[(I-P_{+,n}')Z]\nonumber\\
&=\operatorname{Tr}Z.
\end{align}
Hence $\widehat\Lambda^{(n)}$ is CPTP. Therefore every output attainable in the reduced space optimization can also be attained by a CPTP map on the original reference space. 

Thus, we have proved both directions: for every $n\ge1$ and every finite dimensional output system $E$, the two optimizations have exactly the same attainable output states. Using the Choi formulation of the constrained norm from the previous subsection therefore gives

\begin{equation}
\upsilon_2(\mathcal M_{+,d}^{\otimes n})=\upsilon_2(\mathcal R_{+,d}^{\otimes n}),\quad n\ge1.
\label{eq:norm-suppo_reduc}
\end{equation}
It means the full space $H$ can be replaced by the smaller symmetric input space $\mathsf H_+$ without changing the optimization, even when the post-processing channel acts jointly across all $n$ copies and has an arbitrary finite dimensional output.

\section{ An explicit lower bound}\label{sec:4:lower-bound}

As noted earlier, a lower bound is immediate once we have a chosen channel. Therefore, to obtain an explicitly lower bound on the constrained norm, it is enough to choose one post-processing channel and compute the purity of the resulting output. We know that the fixed Choi state of $\mathcal R_{+,d}$ lives on $A\otimes\mathsf H_+'$, so the post-processing channel must act on the reference space $\mathsf H_+'$. Therefore, we choose on that side the same symmetric partial trace construction used in $\mathcal R_{+,d}$. Let define
\begin{equation}\label{eq:lower-postpr}
\Lambda_\star: \mathcal L(\mathsf H_+')
\longrightarrow \mathcal L(B_1'), \qquad
\Lambda_\star(Z) := \operatorname{Tr}_{B_2'}
(J_+'ZJ_+'^\dagger).
\end{equation}
Thus, $J_+'$ embeds the symmetric reference space into $B_1'\otimes B_2',$ and $\Lambda_\star$ keeps $B_1'$, while discarding $B_2'$. In fact, it is the reference system copy of $\mathcal R_{+,d}$, and therefore CPTP. Applying this particular channel to the reference side of the Choi state gives
\begin{equation}
\omega_\star:=(\mathcal R_{+,d}\otimes\Lambda_\star)
\left(\Phi_{D_d}^{\mathsf H_+\mathsf H_+'}\right).
\label{eq:lower-output-def}
\end{equation}
Since $\Lambda_\star$ is one choice in the optimization, we have
\begin{equation}
\upsilon_2(\mathcal R_{+,d})\ge\|\omega_\star\|_2.
\end{equation}
Then, to obtain a lower bound we need to compute this purity exactly.


We know that both $\mathcal R_{+,d}$ and $\Lambda_\star$ first embed a symmetric space into two $d$ dimensional systems and then discard the second system. It is therefore convenient to apply both embeddings first. Let define
\begin{equation}
|\Psi_\star\rangle:=(J_+\otimes J_+')
|\Phi_{D_d}\rangle_{\mathsf H_+\mathsf H_+'}
\in B_1\otimes B_2\otimes B_1'\otimes B_2',
\label{eq:lower-embedded-vec}
\end{equation}
here four systems appear because
\begin{equation}
\mathsf H_+
\overset{J_+}{\longrightarrow}
B_1\otimes B_2,
\qquad \mathsf H_+' \overset{J_+'}{\longrightarrow}
B_1'\otimes B_2',
\end{equation}
the two channels keep $B_1$ and $B_1'$ and discard $B_2$ and $B_2'$. Therefore
\begin{equation}
\omega_\star=\operatorname{Tr}_{B_2B_2'}
|\Psi_\star\rangle\langle\Psi_\star|.
\label{eq:lower-output-partial-tr}
\end{equation}
By using the maximally entangled vector $\Phi_{D_d}$, in Equation \eqref{eq:Phi-in-basis}, we can write Equation \eqref{eq:lower-embedded-vec} as
\begin{equation}
|\Psi_\star\rangle=\frac1{\sqrt{D_d}}\sum_{a=1}^{D_d}
|e_a\rangle_{B_1B_2} \otimes|e_a\rangle_{B_1'B_2'}.
\end{equation}
For basis indices $i,j,k,\ell$, this gives

\begin{equation}
\begin{aligned}
\langle i,j,k,\ell|\Psi_\star\rangle&=\frac1{\sqrt{D_d}}
\sum_a\langle i,j|e_a\rangle\langle k,\ell|e_a\rangle= \frac1{\sqrt{D_d}}\langle i,j|P_+|k,\ell\rangle.
\end{aligned}
\end{equation}
Since $P_+=(I+\Pi)/2$ we have

\begin{equation}
\langle i,j|P_+|k,\ell\rangle=\frac12 \left(\delta_{ik}\delta_{j\ell}+\delta_{i\ell}\delta_{jk}\right),
\end{equation}
therefore
\begin{equation}
\langle i,j,k,\ell|\Psi_\star\rangle=\frac{
\delta_{ik}\delta_{j\ell}+\delta_{i\ell}\delta_{jk}
}{2\sqrt{D_d}}.
\label{eq:lower-v-coeff}
\end{equation}
This single formula gives every coefficient of $|\Psi_\star\rangle$.


To obtain a matrix entry of $\omega_\star$, we keep the indices belonging to $B_1$ and $B_1'$, and sum over the discarded indices belonging to $B_2$ and $B_2'$
\begin{equation}
\begin{aligned}
\langle i,k|\omega_\star|p,q\rangle
&=\sum_{j,\ell=1}^d
\langle i,j,k,\ell|\Psi_\star\rangle
\langle\Psi_\star|p,j,q,\ell\rangle.
\end{aligned}
\end{equation}
Substituting Equation \eqref{eq:lower-v-coeff}, gives 

\begin{equation}
\begin{aligned}
\langle i,k|\omega_\star|p,q\rangle
&=\frac1{4D_d}\sum_{j,\ell}
\left(\delta_{ik}\delta_{j\ell}
+\delta_{i\ell}\delta_{jk}\right) \times\left(\delta_{pq}\delta_{j\ell}
+\delta_{p\ell}\delta_{jq}\right).
\end{aligned}
\end{equation}
Expanding the four products gives
\begin{equation}
\langle i,k|\omega_\star|p,q\rangle
=\frac{\delta_{ip}\delta_{kq}+(d+2)\delta_{ik}\delta_{pq}
}{4D_d}.\label{eq:lower-output-entriesomegastar}
\end{equation}
The first term is the matrix element of the identity on $B_1\otimes B'_1$. For the second term, recall that

\begin{equation}
\Phi_d^{B_1B'_1} = \frac1d \sum_{i,p=1}^d |i,i\rangle\langle p,p|,
\end{equation}
so $\langle i,k| \Phi_d^{B_1B'_1}|p,q\rangle=(1/d) \delta_{ik}\delta_{pq}.$ Therefore, Equation \eqref{eq:lower-output-entriesomegastar} becomes
\begin{equation}
\omega_\star=\frac{I_{B_1B'_1}+d(d+2)\Phi_d^{B_1B'_1}}{2d(d+1)}.
\label{eq:lower-output-explicit}
\end{equation}
The form in Equation \eqref{eq:lower-output-explicit} makes the spectrum immediate. The operator $\Phi_d^{B_1B'_1}$ is the rank one projector onto $|\Phi_d\rangle$, so the space splits naturally into $\operatorname{span}\{|\Phi_d\rangle\}$ and its orthogonal complement. On $|\Phi_d\rangle$, we have
\begin{equation}
\Phi_d|\Phi_d\rangle=|\Phi_d\rangle,
\end{equation}
so the eigenvalue of $\omega_\star$ is
\begin{equation}
\frac{1+d(d+2)}{2d(d+1)}=\frac{(d+1)^2}{2d(d+1)}
=\frac{d+1}{2d}.
\end{equation}
with multiplicity $1$. On the $d^2-1$ dimensional orthogonal complement we have $\Phi_d=0,$ so the eigenvalue is 
\begin{equation}
 \frac1{2d(d+1)}   
\end{equation}
with multiplicity $d^2-1$. The squared Schatten $2$-norm is therefore given by
\begin{equation}
\begin{aligned}
\|\omega_\star\|_2^2&=\left(\frac{d+1}{2d}\right)^2
+(d^2-1)\left(\frac1{2d(d+1)}\right)^2=\frac{d^2+3d+4}{4d(d+1)}.
\end{aligned}
\end{equation}
For convenient we define 
\begin{equation}
g_d:=\frac{d^2+3d+4}{4d(d+1)}.
\label{eq:gd}
\end{equation}
Since $\Lambda_\star$ is an admissible channel in the Choi formulation of the constrained norm,
\begin{equation}
\upsilon_2(\mathcal R_{+,d})^2\ge g_d.
\end{equation}
Using the exact reduction in Equation \eqref{eq:norm-suppo_reduc}, we know
\begin{equation}
\upsilon_2(\mathcal M_{+,d})^2=\upsilon_2(\mathcal R_{+,d})^2 \ge g_d.
\label{eq:one copy-lower-bound}
\end{equation}
This establishes an achievable value. To prove that it is the optimum, it remains to show that no admissible post-processing channel can produce purity larger than $g_d$.

\section{A matching upper bound}
\label{sec:5:upper-bound}
The previous subsection gave a channel for which the output purity is $g_d$. We now prove that no CPTP post-processing channel can give a larger value. The first step is to rewrite the purity so that the fixed map $\mathcal R_{+,d}$ and the variable channel $\Lambda$ appear separately, because $\mathcal R_{+,d}$ is fixed, while $\Lambda$ is the part we are optimizing over. Once they are separated, we can study the fixed structure once and then prove a bound that holds for every possible $\Lambda$. Let set $E=B'$ here and be an arbitrary finite dimensions output system.
\begin{equation}
\Lambda: \mathcal L(\mathsf H_+')
\longrightarrow \mathcal L(E)
\end{equation}
be any CPTP map, and copies are given by $\widetilde A:=B_1$, $\widetilde E$, and $\widetilde{\mathsf H}_+$ of $A$, $E$, and $\mathsf H_+$, respectively. Recall that for any state $\rho$, we have 
\begin{equation}
\operatorname{Tr}\rho^2=\operatorname{Tr}
\left[(\rho\otimes\rho)\Pi\right],
\end{equation}
where $\Pi$ swaps the two copies. Also, for a map
$\mathcal N$, its adjoint $\mathcal N^\dagger$ is defined by
\begin{equation}
\operatorname{Tr}\left[Y\,\mathcal N(X)\right]=\operatorname{Tr}\left[
\mathcal N^\dagger(Y)\,X\right].
\end{equation}
These standard identities allow us to move the two output swaps backwards through the two channels \cite{Watrous2018TheTheoryofQuantumInformation,Wilde2017QuantumInformationTheory}. Now let define 
\begin{align}
K_d&:=(\mathcal R_{+,d}^\dagger \otimes \mathcal R_{+,d}^\dagger)(\Pi_{A,\widetilde A}),
\label{eq:fixed-swap-kern}
\\
Q_\Lambda&:=(\Lambda^\dagger\otimes\Lambda^\dagger)
(\Pi_{E,\widetilde E}).
\label{eq:chan-swap-ker}
\end{align}
Both operators act on
$\mathsf H_+\otimes\widetilde{\mathsf H}_+$.
The first, $K_d$, depends only on
$\mathcal R_{+,d}$, while all dependence on the post-processing channel is contained in $Q_\Lambda$. Using the swap identity for purity together with
\begin{equation}
\langle\Phi_m| X\otimes Y |\Phi_m\rangle
=\frac1m\operatorname{Tr}(XY^T),
\end{equation}
we obtain
\begin{equation}\label{eq:kernel-objective}
\left\| (\mathcal R_{+,d}\otimes\Lambda)
\left(\Phi_{D_d}^{\mathsf H_+\mathsf H_+'}\right)
\right\|_2^2=\frac1{D_d^2}
\operatorname{Tr}(K_dQ_\Lambda^T),
\end{equation}
where $T$ is the full transpose in the fixed basis. Therefore, the problem is now to bound one trace involving the fixed operator $K_d$ and the variable operator $Q_\Lambda$. For an operator $X$ on $\mathsf H_+\otimes\widetilde{\mathsf H}_+$, we write
\begin{equation}
X_{ab,cf}:=\langle e_a,\widetilde e_b|
X|e_c,\widetilde e_f\rangle.
\end{equation}
and define the partial transpose on the second factor by
\begin{equation}
\mathscr T(X)_{ab,cf}:=X_{af,cb},
\label{eq:partial-transpose}
\end{equation}
and the realignment by
\begin{equation}
\mathscr R(X)_{ab,cf}:=X_{af,bc}.
\label{eq:realignment}
\end{equation}

\begin{lemma}[Properties of the channel kernel]
\label{lem:1channel-k-props}
For every CPTP map
\begin{equation}
\Lambda:
\mathcal L(\mathsf H_+')
\longrightarrow
\mathcal L(E),
\end{equation}
the operator $Q_\Lambda$ defined in
Equation \eqref{eq:chan-swap-ker} satisfies
\begin{equation}
Q_\Lambda=Q_\Lambda^\dagger, \qquad
-I\le Q_\Lambda\le I,
\label{eq:kernel-contraction}
\end{equation}
and
\begin{equation}
\mathscr T(Q_\Lambda)\ge0,
\qquad \mathscr R(Q_\Lambda)\ge0.
\label{eq:kernel-positivity}
\end{equation}
\end{lemma}

\begin{proof}
Since $\Lambda$ is CPTP, $\Lambda^\dagger$ is unital and completely positive. Furthermore, we know $-I \le \Pi_{E,\widetilde E}\le I$, therefore applying $(\Lambda^\dagger\otimes\Lambda^\dagger)$ to $\Pi_{E,\widetilde E}$ gives
\begin{equation}
Q_\Lambda= Q_\Lambda^\dagger, \qquad
-I\le Q_\Lambda\le I.
\end{equation}
Thus, the operator $Q_\Lambda$ itself need not be positive. This proves Equation \eqref{eq:kernel-contraction}.

For the two positivity statements, we choose Kraus operators $L_\alpha$ for $\Lambda$, because it lets us use the complete positivity of $\Lambda$ in an explicit form.
\begin{equation}
\Lambda(X)=\sum_{\alpha} L_\alpha XL_\alpha^\dagger,
\end{equation}
and let
\begin{equation}
|\Omega_E\rangle := \sum_r
|r\rangle_E \otimes
|\widetilde r\rangle_{\widetilde E}.
\end{equation}
Substituting the Kraus representation into the definition of $Q_\Lambda$ and taking the partial transpose gives
\begin{equation}
\mathscr T(Q_\Lambda) =\sum_{\alpha,\beta}
(L_\alpha^\dagger\otimes L_\beta^T)
|\Omega_E\rangle\langle\Omega_E|
(L_\alpha\otimes\overline{L_\beta}),
\label{eq:kernel-pt-kraus}
\end{equation}
where $(L_\beta^T)^\dagger=\overline{L_\beta}$, so each term is positive, and
\begin{equation}
\mathscr T(Q_\Lambda)\ge0.
\end{equation}
For the realignment, now consider the basic matrix units
\begin{equation}
E_{ab}:=|e_a\rangle\langle e_b|,
\end{equation}
definitions of $Q_\Lambda$, Equation \eqref{eq:chan-swap-ker}, and $\mathscr R$, Equation \eqref{eq:realignment}, together with the adjoint relation and the swap identity, give
\begin{equation}
\bigl[\mathscr R(Q_\Lambda)\bigr]_{ab,cf}
=\operatorname{Tr}\left[ \Lambda(E_{ab})^\dagger \Lambda(E_{cf}) \right],
\end{equation}
where each entry is exactly the inner product between the two vectors, therefore $\mathscr R(Q_\Lambda)$ is the Gram matrix of the operators
\begin{equation}
\{\Lambda(E_{ab})\}_{a,b}
\end{equation}
with respect to the Hilbert-Schmidt inner product. Every Gram matrix is positive, hence
\begin{equation}
\mathscr R(Q_\Lambda)\ge0.
\end{equation}
This proves relations \eqref{eq:kernel-positivity}.
\end{proof}
We now turn to the fixed operator $K_d$, Equation \eqref{eq:fixed-swap-kern}. Since it comes from two copies of $\mathcal R_{+,d}$, it is useful to rewrite it on the four underlying $d$ dimensional systems. This will give a simple formula for $K_d$ in terms of permutation operators. More precisely:
\begin{lemma}[The fixed kernel]
\label{lem:fixed-kernel}

Let $P_{(4)}$ be the projector onto the fully symmetric subspace
$\operatorname{Sym}^4(\mathbb C^d)$. Then

\begin{equation}
K_d=\frac{6P_{(4)}-I-\Pi_{\mathsf H_+,\widetilde{\mathsf H}_+}}{4},
\label{eq:k-symmetrizer}
\end{equation}
Moreover,
\begin{equation}
K_d^T=K_d,
\qquad K_d\,\Pi_{\mathsf H_+,\widetilde{\mathsf H}_+}=K_d,
\label{eq:kernel-real-swap}
\end{equation}
and
\begin{equation}
\operatorname{Tr}P_{(4)}=\binom{d+3}{4}
=\frac{d(d+1)(d+2)(d+3)}{24}.
\label{eq:p4_dim}
\end{equation}
\end{lemma}

\begin{proof}
We first rewrite $K_d$ using the definition of
$\mathcal R_{+,d}$. The Hilbert-Schmidt adjoint of Equation \eqref{eq:sym-partial-tr} gives
\begin{equation}\label{eq:r-adjoint}
\mathcal R_{+,d}^\dagger(Y)=J_+^\dagger
(Y\otimes I_{B_2})J_+.
\end{equation}
To apply this to the two copies $\mathsf H_+\otimes\widetilde{\mathsf H}_+$, in $K_d$, consider this as the subspace of four $d$ dimensional systems
\begin{equation}
1:=B_1,\qquad
2:=B_2,\qquad
3:=\widetilde B_1,\qquad
4:=\widetilde B_2.
\end{equation}
The first pair $(1,2)$ lies in its symmetric subspace, and the second pair $(3,4)$ does the same. Therefore the projector onto this space is
\begin{equation}
P:=P_+^{12}P_+^{34}.
\end{equation}
The swap $\Pi_{A,\widetilde A}$ in the definition of $K_d$ exchanges systems $1$ and $3$. Using Equation \eqref{eq:r-adjoint} on both copies therefore gives
\begin{equation}
K_d=\left.P\Pi_{13}P\right|_{\operatorname{ran}P}.
\label{eq:kernel-four-syst}
\end{equation}
We now compare this operator with the fully symmetric projector $P_{(4)}$.

The projector $P_{(4)}$ is the average of all $4!=24$ permutations of the four systems. Because we are already restricted to $\operatorname{ran}P$, exchanging $1$ with $2$, or $3$ with $4$, has no effect. After this restriction, the $24$ permutations fall into three groups. 
\begin{enumerate}
 \item  Four permutations preserve the two pairs and reduce to
\begin{equation}
I.
\end{equation}
 \item  Four permutations exchange the two pairs and reduce to $\Pi_{\mathsf H_+,\widetilde{\mathsf H}_+},$ which, in the four system picture, is the restriction of
\begin{equation}
\Pi_{13}\Pi_{24}.
\end{equation}
\item The remaining sixteen permutations exchange one member of each pair and reduce to the same compressed operator appearing in Equation \eqref{eq:kernel-four-syst}, namely $K_d$.
\end{enumerate}
Therefore
\begin{equation}
P_{(4)} =\frac{4I+4\Pi_{\mathsf H_+,\widetilde{\mathsf H}_+}+ 16K_d}{24},
\end{equation}
and
\begin{equation}
K_d = \frac{ 6P_{(4)} -I-\Pi_{\mathsf H_+,\widetilde{\mathsf H}_+}}{4}.
\end{equation}
This proves Equation \eqref{eq:k-symmetrizer}.
Moreover, the operators $P_{(4)}$, $I$, and $\Pi_{\mathsf H_+,\widetilde{\mathsf H}_+}$ are all real in the fixed basis, therefore
\begin{equation}\label{eq:k_d-tras}
K_d^T=K_d.
\end{equation}
Also, a fully symmetric vector is unchanged when the two pairs are exchanged, so
\begin{equation}
P_{(4)}\Pi_{\mathsf H_+,\widetilde{\mathsf H}_+}=P_{(4)}.
\end{equation}
Multiplying Equation \eqref{eq:k-symmetrizer} on the right by $\Pi_{\mathsf H_+,\widetilde{\mathsf H}_+}$ and considering
$\Pi_{\mathsf H_+,\widetilde{\mathsf H}_+}^2=I$ gives
\begin{equation}
\begin{aligned}
K_d\Pi_{\mathsf H_+,\widetilde{\mathsf H}_+}
&=\frac{6P_{(4)}\Pi_{\mathsf H_+,\widetilde{\mathsf H}_+}
-\Pi_{\mathsf H_+,\widetilde{\mathsf H}_+}
-\Pi_{\mathsf H_+,\widetilde{\mathsf H}_+}^2}{4}\\
&=\frac{6P_{(4)}-\Pi_{\mathsf H_+,\widetilde{\mathsf H}_+}
-I}{4}\\
&=K_d.
\end{aligned}
\end{equation}
This proves Equation \eqref{eq:kernel-real-swap}.
Finally, the dimension of the fourth symmetric power of a
$d$ dimension space is
\begin{equation}
\dim\operatorname{Sym}^4(\mathbb C^d)=\binom{d+3}{4},
\end{equation}
so we obtain
\begin{equation}
\operatorname{Tr}P_{(4)}=\binom{d+3}{4}=\frac{
d(d+1)(d+2)(d+3)}{24}.
\end{equation}
\end{proof}
The first consequence of this Lemma is that the transpose in Equation \eqref{eq:kernel-objective} disappears. we have $K_d^T=K_d$, by Equation \eqref{eq:k_d-tras} we have
\begin{equation}
\operatorname{Tr}(K_dQ_\Lambda^T)
=\operatorname{Tr}(K_d^TQ_\Lambda)
=\operatorname{Tr}(K_dQ_\Lambda).
\end{equation}
Therefore
\begin{equation}
\left\|(\mathcal R_{+,d}\otimes\Lambda)
\left(\Phi_{D_d}^{\mathsf H_+\mathsf H_+'}\right)
\right\|_2^2 = \frac1{D_d^2}
\operatorname{Tr}(K_dQ_\Lambda).
\label{eq:kernel-objective-real}
\end{equation}
Now for finding the upper bound we want to bound
\begin{equation}
\operatorname{Tr}(K_dQ_\Lambda),
\end{equation}
however, $Q_\Lambda$ itself need not to be positive. Therefore a usual operator inequality such as $K_d\le\mathsf L_d$ would not be enough. However, we already know 
$\mathscr T(Q_\Lambda)$ and $\mathscr R(Q_\Lambda)$ are positive. This suggests looking for positive operators $Z_d$ and $\mathsf L_d$, together with $\eta_d>0$, such that

\begin{equation}
\mathsf L_d-K_d=\eta_d\left(\mathscr T(Z_d)+\mathscr R^\dagger(Z_d)\right),
\label{eq:upper}
\end{equation}
where $\mathscr R^\dagger$ is the Hilbert-Schmidt adjoint of the realignment map. If such operators exist, then
\begin{equation}
\begin{aligned}
\operatorname{Tr}[(\mathsf L_d-K_d)Q_\Lambda]
&=\eta_d\left(\operatorname{Tr}
[Z_d\mathscr T(Q_\Lambda)]+\operatorname{Tr}
[Z_d\mathscr R(Q_\Lambda)]
\right)\\
&\ge0.
\end{aligned}
\end{equation}
Thus
\begin{equation}
\operatorname{Tr}(K_dQ_\Lambda) \le
\operatorname{Tr}(\mathsf L_dQ_\Lambda).
\end{equation}
If also $\mathsf L_d\ge0$, then $Q_\Lambda\le I$ gives
\begin{equation}
\operatorname{Tr}(\mathsf L_dQ_\Lambda)
\le \operatorname{Tr}\mathsf L_d.
\end{equation}
So it remains to construct $Z_d$ and $\mathsf L_d$.

\begin{lemma}[Construction of $Z_d$]
\label{lem:positive-Z}
Define
\begin{equation}
Z_d = I - \frac4{d+2}\mathscr R(K_d)
+ \frac{2}{(d+1)(d+2)} |\Omega_+\rangle\langle\Omega_+|,
\label{eq:positive-test-op}
\end{equation}
where
\begin{equation}
|\Omega_+\rangle := \sum_{a=1}^{D_d}
|e_a\rangle \otimes |\widetilde e_a\rangle.
\end{equation}
Then
\begin{equation}
Z_d\ge0.
\end{equation}
More precisely, $Z_d$ is the orthogonal projector onto the part corresponding to
$\ker\mathcal R_{+,d}$.
\end{lemma}

\begin{proof}
Let $E_{ab}=|e_a\rangle\langle e_b|.$ From the definitions of $K_d$ and $\mathscr R$, we have
\begin{equation}
\langle e_a,\widetilde e_b| \mathscr R(K_d)
|e_c,\widetilde e_f\rangle = \left\langle
E_{ab}, \mathcal R_{+,d}^\dagger \mathcal R_{+,d}(E_{cf})
\right\rangle_{\mathrm{HS}}.
\label{eq:realigned-k-action}
\end{equation}
Thus, in the orthonormal operator basis $\{E_{ab}\}$, the matrix $\mathscr R(K_d)$ represents the map
\begin{equation}
\mathcal R_{+,d}^\dagger
\mathcal R_{+,d},
\end{equation}
on the Hilbert-Schmidt operator space $\mathcal L(\mathsf H_+)$. We therefore first determine the eigenvalues of this map. By Equation \eqref{eq:r-adjoint}, we can write
\begin{equation}
\mathcal R_{+,d} \bigl( \mathcal R_{+,d}^\dagger(Y)
\bigr) = \operatorname{Tr}_{B_2} \left[P_+
(Y\otimes I) P_+\right].
\end{equation}
Substituting $P_+=\frac{I+\Pi}{2}$ gives
\begin{align}
&\operatorname{Tr}_{B_2}
\left[ P_+(Y\otimes I)P_+\right]=\nonumber\\
& \frac14 \operatorname{Tr}_{B_2} \Bigl[ (Y\otimes I) + (Y\otimes I)\Pi + \Pi(Y\otimes I)
+ \Pi(Y\otimes I)\Pi \Bigr].
\end{align}
The four partial traces are
\begin{equation}
dY,\qquad Y,\qquad Y,\qquad (\operatorname{Tr}Y)I,
\end{equation}
so we obtain 
\begin{equation}
\mathcal R_{+,d} \bigl(\mathcal R_{+,d}^\dagger(Y)
\bigr) = \frac{d+2}{4}Y + \frac{\operatorname{Tr}Y}{4}I_A.
\label{eq:r-radjoint}
\end{equation}
From this easily we can see
\begin{equation}
\mathcal R_{+,d}^\dagger(I_A)
=I_{\mathsf H_+}, \qquad \text{ and } \qquad 
\mathcal R_{+,d}(I_{\mathsf H_+})=\frac{d+1}{2}I_A.
\end{equation}
Equation \eqref{eq:r-radjoint} is invertible as a map on
$\mathcal L(A)$. Therefore
\begin{equation}
\operatorname{ran}\mathcal R_{+,d}^\dagger
=(\ker\mathcal R_{+,d})^\perp.
\end{equation}
Separating the identity direction from the traceless directions gives the Hilbert-Schmidt orthogonal decomposition
\begin{equation}
\begin{aligned}
\mathcal L(\mathsf H_+)
={}& \operatorname{span}\{I_{\mathsf H_+}\} \\
&\oplus \left\{ \mathcal R_{+,d}^\dagger(Y):
\operatorname{Tr}Y=0 \right\} \\
&\oplus \ker\mathcal R_{+,d}.
\end{aligned}
\label{eq:op-sp-decompos}
\end{equation}
On these three subspaces, $\mathcal R_{+,d}^\dagger\mathcal R_{+,d}$ has eigenvalues
\begin{equation}
\frac{d+1}{2}, \qquad \frac{d+2}{4}, \qquad 0,
\end{equation}
respectively. Now we construct an operator that vanishes on the first two subspaces and equals the identity on the third. Let start with
\begin{equation}
I-\alpha\,\mathscr R(K_d).
\end{equation}
On the traceless part, the eigenvalue is
\begin{equation}
1-\alpha\frac{d+2}{4}.
\end{equation}
We want this to be zero, therefore
\begin{equation}
\alpha = \frac4{d+2}.
\end{equation}
With this choice, the eigenvalue on the identity direction is
\begin{equation}
1 - \frac4{d+2} \frac{d+1}{2} =-\frac d{d+2}.
\end{equation}
We therefore need one more term that acts only on the identity direction. The operator
\begin{equation}
|\Omega_+\rangle\langle\Omega_+|
\end{equation}
does exactly this: it acts with eigenvalue $D_d$ on the identity direction and vanishes on the two orthogonal subspaces. Therefore we choose $\beta$ so that
\begin{equation}
-\frac d{d+2} + \beta D_d = 0.
\end{equation}
Recall $D_d=\frac{d(d+1)}2,$ we obtain
\begin{equation}
\beta=\frac{2}{(d+1)(d+2)}.
\end{equation}
Therefore
\begin{equation}
Z_d = I - \frac4{d+2}\mathscr R(K_d) +
\frac{2}{(d+1)(d+2)} |\Omega_+\rangle\langle\Omega_+|.
\end{equation}
Its eigenvalues on the three subspaces in
Equation \eqref{eq:op-sp-decompos} are
\begin{equation}
0,\qquad0,\qquad1.
\end{equation}
Therefore
\begin{equation}
Z_d\ge0.
\end{equation}
In fact, $Z_d$ is exactly the orthogonal projector onto the part corresponding to
$\ker\mathcal R_{+,d}$.
\end{proof}

\begin{lemma}[Construction of the comparison operator]
\label{lem:4comparison_op}

Let $Z_d$ be the positive operator constructed in
Lemma~\ref{lem:positive-Z}. Define
\begin{equation}
\eta_d := \frac{d+1}{4(d+3)},
\label{eq:eta-d}
\end{equation}
and
\begin{equation}
\ell_d:= \frac{3(d^2+3d+4)} {2(d+2)(d+3)}, \qquad
\mathsf L_d := \ell_d P_{(4)}.
\label{eq:Ld}
\end{equation}
Then
\begin{equation}
\mathsf L_d\ge0,
\end{equation}
and
\begin{equation}\label{eq:upper-L-K}
\mathsf L_d-K_d = \eta_d \left( \mathscr T(Z_d) + \mathscr R^\dagger(Z_d) \right).
\end{equation}
\end{lemma}

\begin{proof}
We start from the expression for $Z_d$, Equation \eqref{eq:positive-test-op} and apply $\mathscr T$ and $\mathscr R^\dagger$ to its three terms.
From the definitions of partial transpose Equation \eqref{eq:partial-transpose} and realignment Equation \eqref{eq:realignment}, we have
\begin{equation}
\mathscr R(X)=\mathscr T(X\Pi_{\mathsf H_+,\widetilde{\mathsf H}_+}),
\qquad \mathscr R^\dagger(X)=\mathscr T(X)\Pi_{\mathsf H_+,\widetilde{\mathsf H}_+}.
\label{eq:realignment-transpose-relation}
\end{equation}
Since, by Lemma \ref{lem:fixed-kernel}, we know
\begin{equation}
K_d\Pi_{\mathsf H_+,\widetilde{\mathsf H}_+}=K_d,
\end{equation}
tehrefore, we obtain
\begin{equation}
\mathscr T(\mathscr R(K_d)) = K_d, \qquad \mathscr R^\dagger(\mathscr R(K_d)) = K_d.
\end{equation}
For the identity and the ranked one operator
$|\Omega_+\rangle\langle\Omega_+|$, the definitions give
\begin{equation}
\mathscr T(I)=I,
\qquad
\mathscr R^\dagger(I)=\Pi_{\mathsf H_+,\widetilde{\mathsf H}_+},
\end{equation}
and
\begin{equation}
\mathscr T \left( |\Omega_+\rangle\langle\Omega_+|
\right)= \Pi_{\mathsf H_+,\widetilde{\mathsf H}_+},
\qquad \mathscr R^\dagger
\left( |\Omega_+\rangle\langle\Omega_+| \right) = I.
\end{equation}
Therefore, applying these identities to
Equation \eqref{eq:positive-test-op} gives
\begin{align}
\mathscr T(Z_d) &=I-\frac4{d+2}K_d+\frac{2}{(d+1)(d+2)}
\Pi_{\mathsf H_+,\widetilde{\mathsf H}_+},
\label{eq:transformed-test-pt}\\
\mathscr R^\dagger(Z_d) &= \Pi_{\mathsf H_+,\widetilde{\mathsf H}_+} -\frac4{d+2}K_d
+ \frac{2}{(d+1)(d+2)} I.
\label{eq:transformed-test-realignment}
\end{align}
Adding these two equations gives
\begin{equation}
\begin{aligned}
\mathscr T(Z_d) + \mathscr R^\dagger(Z_d)&=
\left(1+\frac{2}{(d+1)(d+2)}\right)
(I+\Pi_{\mathsf H_+,\widetilde{\mathsf H}_+}) -\frac8{d+2}K_d.
\end{aligned}
\end{equation}
From Equation \eqref{eq:k-symmetrizer}, we know
\begin{equation}
I+\Pi_{\mathsf H_+,\widetilde{\mathsf H}_+}
=6P_{(4)}-4K_d.
\end{equation}
Substituting this gives
\begin{align}
\mathscr T(Z_d)+
\mathscr R^\dagger(Z_d)
&= 6 \left[ 1+\frac{2}{(d+1)(d+2)} \right] P_{(4)}\nonumber\\
&\quad- \left[4 \left( 1+\frac{2}{(d+1)(d+2)} \right)
+ \frac8{d+2} \right] K_d.
\end{align}
The two coefficients simplify to
\begin{equation}
6 \left[ 1+\frac{2}{(d+1)(d+2)} \right] =
\frac{ 6(d^2+3d+4)}{(d+1)(d+2)},
\end{equation}
and
\begin{equation}
4\left[1+\frac{2}{(d+1)(d+2)}\right]+\frac8{d+2}=
\frac{4(d+3)}{d+1}.
\end{equation}
Therefore we obtain
\begin{equation}
\mathscr T(Z_d)+\mathscr R^\dagger(Z_d)
=\frac{ 6(d^2+3d+4)}{(d+1)(d+2)}P_{(4)}-\frac{4(d+3)}{d+1}
K_d.
\label{eq:transformed-test-sum}
\end{equation}
We want the coefficient of $K_d$ to become $-1$. This fixes the scaling factor as $\eta_d \frac{4(d+3)}{d+1}=1,$
so
\begin{equation}
\eta_d=\frac{d+1}{4(d+3)}.
\end{equation}
Multiplying Equation \eqref{eq:transformed-test-sum} by $\eta_d$ gives
\begin{equation}
\eta_d \left(\mathscr T(Z_d)+\mathscr R^\dagger(Z_d)
\right)=\frac{3(d^2+3d+4)}{2(d+2)(d+3)}P_{(4)}-K_d.
\end{equation}
This determines
\begin{equation}
\ell_d =\frac{3(d^2+3d+4)}{2(d+2)(d+3)}
\end{equation}
and hence
\begin{equation}
\mathsf L_d=\ell_dP_{(4)}.
\end{equation}
Therefore
\begin{equation}
\mathsf L_d-K_d=\eta_d \left(\mathscr T(Z_d)+\mathscr R^\dagger(Z_d)\right).
\end{equation}
Finally, $\ell_d>0$ and $P_{(4)}$ is a projection, so
\begin{equation}
\mathsf L_d\ge0.
\end{equation}
\end{proof}
\begin{theorem}[Exact one copy constrained norm]
\label{thm:sw-one copy-norm}
For every integer $d\ge2$,
\begin{equation}
\upsilon_2(\mathcal M_{+,d})=\upsilon_2(\mathcal R_{+,d})=\sqrt{\frac{d^2+3d+4}{4d(d+1)}}.
\label{eq:exact-one copy-norm}
\end{equation}
The value for $\mathcal R_{+,d}$ is attained by the channel $\Lambda_\star$ from Equation \eqref{eq:lower-postpr}.
\end{theorem}

\begin{proof}
Let $\Lambda: \mathcal L(\mathsf H_+')\longrightarrow
\mathcal L(E)$ be any CPTP post-processing map. From Lemma~\ref{lem:4comparison_op}. Taking the trace against $Q_\Lambda$, in Equation \eqref{eq:upper-L-K} gives
\begin{equation}
\begin{aligned}
\operatorname{Tr}[(\mathsf L_d-K_d)Q_\Lambda]
&= \eta_d \left( \operatorname{Tr} [Z_d\mathscr T(Q_\Lambda)] + \operatorname{Tr}
[Z_d\mathscr R(Q_\Lambda)] \right) \ge0,
\end{aligned}
\end{equation}
because $Z_d\ge0$ by Lemma~\ref{lem:positive-Z} and by Lemma~\ref{lem:channel-k-props} we have $\mathscr T(Q_\Lambda)\ge0$, and $\mathscr R(Q_\Lambda)\ge0$. Therefore
\begin{equation}
\operatorname{Tr}(K_dQ_\Lambda)
\le \operatorname{Tr}(\mathsf L_dQ_\Lambda).
\end{equation}
Since again by Lemma \ref{lem:channel-k-props}, we know $Q_\Lambda\le I$ and by Lemma \ref{lem:4comparison_op} $\mathsf L_d\ge0,$ we also have
\begin{equation}
\operatorname{Tr}(\mathsf L_dQ_\Lambda)
\le \operatorname{Tr}\mathsf L_d.
\end{equation}
Now again by Lemme \ref{lem:4comparison_op}, $\mathsf L_d=\ell_dP_{(4)},$ and using Equation \eqref{eq:p4_dim}, we obtain
\begin{equation}\label{ew:TR-L_d}
\begin{aligned}
\operatorname{Tr}\mathsf L_d
&= \ell_d\operatorname{Tr}P_{(4)}\\
&=\frac{3(d^2+3d+4)}{2(d+2)(d+3)}\frac{d(d+1)(d+2)(d+3)}{24
}\\
&=\frac{d(d+1)(d^2+3d+4)}{16}\\
&= D_d^2g_d.
\end{aligned}
\end{equation}
Therefore Equation \eqref{eq:kernel-objective-real} gives
\begin{equation}
\left\| (\mathcal R_{+,d}\otimes\Lambda)
\left(\Phi_{D_d}^{\mathsf H_+\mathsf H_+'}\right)
\right\|_2^2 \le g_d.
\end{equation}
Since this holds for every $E$ and every CPTP map $\Lambda:
\mathcal L(\mathsf H_+') \longrightarrow \mathcal L(E)$, we can deduce
\begin{equation}
\upsilon_2(\mathcal R_{+,d})^2 \le g_d.
\end{equation}
The lower bound from Equation \eqref{eq:one copy-lower-bound} gives $\upsilon_2(\mathcal R_{+,d})^2 \ge g_d,$
so
\begin{equation}
\upsilon_2(\mathcal R_{+,d})^2= g_d.
\end{equation}
Finally, Equation \eqref{eq:norm-suppo_reduc} gives
\begin{equation}
\upsilon_2(\mathcal M_{+,d})=\upsilon_2(\mathcal R_{+,d})=
\sqrt{\frac{d^2+3d+4}{4d(d+1)}},
\end{equation}
which proves the result.
\end{proof}


\section{Multiplicativity under tensor powers}
\label{sec:6:tensor-multipl}

We now pass from one copy to $n$ copies. For the reduced map, the quantity to determine is
\begin{equation}
\upsilon_2(\mathcal R_{+,d}^{\otimes n}).
\end{equation}
By the channel formulation of the constrained norm, the optimization allows an arbitrary CPTP map
\begin{equation}
\Lambda^{(n)}:\mathcal L((\mathsf H_+')^{\otimes n})
\longrightarrow \mathcal L(E),
\end{equation}
where $E$ is any finite dimensional system. The superscript $(n)$ only indicates that the channel acts on the full $n$ copy reference system; no product form is assumed.

The lower bound is already clear from the one copy result. We may choose
\begin{equation}
\Lambda^{(n)}=\Lambda_\star^{\otimes n}.
\end{equation}
Then the output is $\omega_\star^{\otimes n},$ and therefore
\begin{equation}
\|\omega_\star^{\otimes n}\|_2^2=\bigl(\|\omega_\star\|_2^2\bigr)^n
=g_d^n,
\end{equation}
thus, we obtain
\begin{equation}
\upsilon_2(\mathcal R_{+,d}^{\otimes n})^2
\ge g_d^n.
\label{eq:tensor-lower-bound}
\end{equation}
The only remaining question is whether a joint channel
$\Lambda^{(n)}$, which may correlate all $n$ copies, can give a larger value. To study this, we use the same kernel description as in the one copy case, and define
\begin{equation}
Q_{\Lambda^{(n)}}:=\left((\Lambda^{(n)})^\dagger \otimes
(\Lambda^{(n)})^\dagger \right)(\Pi_{E,\widetilde E}).
\label{eq:joint-channel-kernel}
\end{equation}
We treat $(\mathsf H_+')^{\otimes n}$ as one input system. Therefore, Lemma \ref{lem:1channel-k-props} applies exactly as before. In particular,
\begin{equation}
Q_{\Lambda^{(n)}}\le I,
\end{equation}
and the corresponding partial transpose and realignment are positive. To write these operations copy by copy, use the ordering
\begin{equation}
\mathsf H_+^{\otimes n}
\otimes \widetilde{\mathsf H}_+^{\otimes n}
\simeq \bigotimes_{k=1}^n \left( \mathsf H_{+,k} \otimes
\widetilde{\mathsf H}_{+,k} \right),
\label{eq:doubled_tensor-order}
\end{equation}
and define
\begin{equation}
\mathscr T_n := \mathscr T^{\otimes n},
\qquad \mathscr R_n := \mathscr R^{\otimes n}.
\label{eq:tensor-transforms}
\end{equation}
Thus
\begin{equation}
Q_{\Lambda^{(n)}}\le I, \qquad \mathscr T_n(Q_{\Lambda^{(n)}})\ge0,
\qquad \mathscr R_n(Q_{\Lambda^{(n)}})\ge0.
\label{eq:joint-kern_prop}
\end{equation}
The fixed kernel also tensorizes. Since the output swap on
$A^{\otimes n}\otimes\widetilde A^{\otimes n}$ is the tensor product of the $n$ one copy swaps, the fixed kernel is
\begin{equation}
K_d^{\otimes n}.
\end{equation}
Therefore the same purity calculation used for one copy gives
\begin{equation}
\left\| (\mathcal R_{+,d}^{\otimes n}\otimes\Lambda^{(n)})
(\Phi_{D_d}^{\otimes n}) \right\|_2^2 = \frac1{D_d^{2n}}
\operatorname{Tr} \left[ K_d^{\otimes n} Q_{\Lambda^{(n)}}
\right],
\label{eq:tensor-kern objec}
\end{equation}
where we used $K_d^T=K_d$. Therefore, the $n$ copy problem has become the following: we need to prove
\begin{equation}
\operatorname{Tr}
\left[ K_d^{\otimes n} Q_{\Lambda^{(n)}} \right]
\le D_d^{2n}g_d^n
\end{equation}
for every joint channel $\Lambda^{(n)}$.

From the last section for one copy, this bound came from the comparison
\begin{equation}
\mathsf L_d-K_d = \eta_d \left( \mathscr T(Z_d)+\mathscr R^\dagger(Z_d) \right).
\end{equation}
So for $n$ copies this is natural to compare $\mathsf L_d^{\otimes n}$ and $K_d^{\otimes n}$. By adding and subtracting the intermediate tensor products, we obtain the elementary telescoping identity
\begin{equation}
\mathsf L_d^{\otimes n} - K_d^{\otimes n}=\sum_{k=1}^n
\mathsf L_d^{\otimes(k-1)} \otimes
(\mathsf L_d-K_d) \otimes K_d^{\otimes(n-k)}.
\label{eq:tensor-telescoping-identity}
\end{equation}
This formula tells us how to build the $n$ copy comparison from the one copy relation: in the $k$-th term, $Z_d$ should reproduce $\mathsf L_d-K_d$, while the factors before it should reproduce $\mathsf L_d$ and the factors after it should reproduce $K_d$.

To do this while keeping the resulting $n$ copy operator positive, we need positive operators that are sent back to
$\mathsf L_d$ and $K_d$ by both transformations. We verify this next.


\begin{lemma}[Positive factors for the tensor construction]
\label{lem:tensor.positive.factor}

For $X\in\{K_d,\mathsf L_d\}$,

\begin{equation}
\mathscr T(X)=\mathscr R(X)\ge0,
\label{eq:crossed-positive}
\end{equation}
and
\begin{equation}
\mathscr T\bigl(\mathscr T(X)\bigr)
=\mathscr R^\dagger\bigl(\mathscr T(X)\bigr)=X.
\label{eq:common-preimages}
\end{equation}
\end{lemma}

\begin{proof}
We first consider $K_d$. From Lemma~\ref{lem:fixed-kernel}, we know $K_d\Pi_{\mathsf H_+,\widetilde{\mathsf H}_+}=K_d.$ Using the relation \eqref{eq:realignment-transpose-relation} we obtain
\begin{equation}
\mathscr R(K_d)=\mathscr T(K_d).
\end{equation}
The operator $K_d$ is the channel kernel associated with
$\mathcal R_{+,d}$. Therefore the realignment positivity proved in Lemma \ref{lem:channel-k-props}, now applied to $\mathcal R_{+,d}$, gives
\begin{equation}
\mathscr R(K_d)\ge0.
\end{equation}
Therefore
\begin{equation}
\mathscr T(K_d)=\mathscr R(K_d)\ge0.
\label{eq:crossed-kernel-positive}
\end{equation}
We now consider $\mathsf L_d$. Recall from Equation \eqref{eq:Ld} that $\mathsf L_d=\ell_dP_{(4)}$ for $\ell_d>0$. Since $P_{(4)}$ projects onto the fully symmetric subspace, exchanging the two pairs leaves it unchanged: $P_{(4)}\Pi_{\mathsf H_+,\widetilde{\mathsf H}_+}=P_{(4)}.$ Therefore
\begin{equation}
\mathscr T(P_{(4)})=\mathscr R(P_{(4)}).
\end{equation}
Now we show that this operator is positive. From Equation \eqref{eq:k-symmetrizer}, we have
\begin{equation}
6P_{(4)}=I+\Pi_{\mathsf H_+,\widetilde{\mathsf H}_+}+4K_d.
\end{equation}
Applying $\mathscr R$ gives
\begin{equation}
6\mathscr R(P_{(4)})=\mathscr R(I)+\mathscr R(\Pi_{\mathsf H_+,\widetilde{\mathsf H}_+})+4\mathscr R(K_d).
\end{equation}
From the definition of realignment \eqref{eq:realignment},
\begin{equation}
\mathscr R(I) =|\Omega_+\rangle\langle\Omega_+|,
\qquad \mathscr R(\Pi_{\mathsf H_+,\widetilde{\mathsf H}_+}) =I.
\end{equation}
Thus
\begin{equation}
\mathscr R(P_{(4)})=\frac{|\Omega_+\rangle\langle\Omega_+
|+I+4\mathscr R(K_d)}{6}.
\label{eq:realigned-p4}
\end{equation}
All three terms on the right are positive. Therefore
\begin{equation}
\mathscr R(P_{(4)})\ge0.
\end{equation}
Since $\ell_d>0$,
\begin{equation}
\mathscr R(\mathsf L_d) = \ell_d\mathscr R(P_{(4)}) \ge0.
\end{equation}
Together with
$\mathscr T(P_{(4)})=\mathscr R(P_{(4)})$, this gives

\begin{equation}
\mathscr T(\mathsf L_d) =\mathscr R(\mathsf L_d) \ge0.
\label{eq:crossed.compar.positive}
\end{equation}
We finally check that these positive operators are mapped back to the original ones.

Partial transpose on the second factor is an involution, so
$\mathscr T(\mathscr T(X))=X.$ Also, from the relation between realignment and partial transpose,
\begin{equation}
\mathscr R^\dagger(Y)
=\mathscr T(Y)\Pi_{\mathsf H_+,\widetilde{\mathsf H}_+}.
\end{equation}
Therefore
\begin{equation}
\mathscr R^\dagger(\mathscr T(X))
=X\Pi_{\mathsf H_+,\widetilde{\mathsf H}_+}.
\end{equation}
Now for $X=K_d$, we obtain
\begin{equation}
K_d\Pi_{\mathsf H_+,\widetilde{\mathsf H}_+}=K_d,
\end{equation}
and for $X=\mathsf L_d=\ell_dP_{(4)}$,
\begin{equation}
\mathsf L_d\Pi_{\mathsf H_+,\widetilde{\mathsf H}_+}
=\mathsf L_d.
\end{equation}
Therefore, for both choices,
\begin{equation}
\mathscr R^\dagger(\mathscr T(X))=X.
\end{equation}
This proves Equation \eqref{eq:common-preimages}.
\end{proof}

We now have the factors needed in each term of the tensor expansion. The middle factor $Z_d$ is positive and produces
$\mathsf L_d-K_d$, while
$\mathscr T(\mathsf L_d)$ and
$\mathscr T(K_d)$ are positive and are mapped back to
$\mathsf L_d$ and $K_d$, respectively. We can therefore follow the telescoping expansion term by term.

\begin{lemma}[Tensor power comparison]
\label{lem:tensor-comparison}

For every $n\ge1$, define
\begin{equation}
Z_{d,n} := \sum_{k=1}^n
\bigl(\mathscr T(\mathsf L_d)\bigr)^{\otimes(k-1)}
\otimes Z_d \otimes
\bigl(\mathscr T(K_d)\bigr)^{\otimes(n-k)},
\label{eq:tensor-positive-op}
\end{equation}
where factors with tensor power $0$ are omitted. Then
\begin{equation}
Z_{d,n}\ge0,
\end{equation}
and
\begin{equation}\label{eq:tensor-identity}
\mathsf L_d^{\otimes n} - K_d^{\otimes n}
= \eta_d \left( \mathscr T_n(Z_{d,n}) + \mathscr R_n^\dagger(Z_{d,n}) \right).
\end{equation}
\end{lemma}

\begin{proof}
Consider the $k$-th term in
Equation \eqref{eq:tensor-positive-op}:
\begin{equation}
\bigl(\mathscr T(\mathsf L_d)\bigr)^{\otimes(k-1)}
\otimes Z_d \otimes \bigl(\mathscr T(K_d)\bigr)^{\otimes(n-k)}.
\end{equation}
Every factor is positive by
Lemma \ref{lem:tensor.positive.factor} and
Lemma \ref{lem:positive-Z}. Therefore every tensor product in the sum is positive, and therefore
\begin{equation}
Z_{d,n}\ge0.
\end{equation}
Now apply $\mathscr T_n=\mathscr T^{\otimes n}$. By
Equation \eqref{eq:common-preimages},
\begin{equation}
\mathscr T(\mathscr T(\mathsf L_d))
= \mathsf L_d, \qquad \mathscr T(\mathscr T(K_d)) = K_d.
\end{equation}
Therefore
\begin{equation}
\begin{aligned}
\mathscr T_n(Z_{d,n}) = \sum_{k=1}^n \mathsf L_d^{\otimes(k-1)} \otimes \mathscr T(Z_d) \otimes K_d^{\otimes(n-k)}.
\end{aligned}
\end{equation}
Similarly,
\begin{equation}
\mathscr R^\dagger(\mathscr T(\mathsf L_d))
= \mathsf L_d, \qquad
\mathscr R^\dagger(\mathscr T(K_d)) =K_d,
\end{equation}
so
\begin{equation}
\begin{aligned}
\mathscr R_n^\dagger(Z_{d,n})
= \sum_{k=1}^n \mathsf L_d^{\otimes(k-1)}
\otimes \mathscr R^\dagger(Z_d)
\otimes K_d^{\otimes(n-k)}.
\end{aligned}
\end{equation}
Adding the two equations and multiplying by $\eta_d$, gives
\begin{equation}
\begin{aligned}
&\eta_d \left( \mathscr T_n(Z_{d,n}) +
\mathscr R_n^\dagger(Z_{d,n}) \right)\\
&= \sum_{k=1}^n \mathsf L_d^{\otimes(k-1)} \otimes \eta_d \left( \mathscr T(Z_d) + \mathscr R^\dagger(Z_d) \right) \otimes K_d^{\otimes(n-k)}.
\end{aligned}
\end{equation}
The one copy relation
Equation \eqref{eq:upper} gives
\begin{equation}
\eta_d\left(\mathscr T(Z_d)+\mathscr R^\dagger(Z_d)\right)=\mathsf L_d-K_d.
\end{equation}
Therefore
\begin{equation}
\begin{aligned}
&\eta_d\left(\mathscr T_n(Z_{d,n})+\mathscr R_n^\dagger(Z_{d,n})\right)\\
&= \sum_{k=1}^n \mathsf L_d^{\otimes(k-1)}
\otimes (\mathsf L_d-K_d) \otimes K_d^{\otimes(n-k)}.
\end{aligned}
\end{equation}
By the telescoping identity established above, the right hand side is
\begin{equation}
\mathsf L_d^{\otimes n}-K_d^{\otimes n}.
\end{equation}
This proves Equation \eqref{eq:tensor-identity}.
\end{proof}
\begin{theorem}[Multiplicativity under tensor powers]
\label{thm:tensor-multipli}

For every integer $d\ge2$ and every integer $n\ge1$,
\begin{equation}
\begin{aligned}
\upsilon_2(\mathcal M_{+,d}^{\otimes n})^2
&= \upsilon_2(\mathcal R_{+,d}^{\otimes n})^2\\
&= g_d^n = \left( \frac{d^2+3d+4}{4d(d+1)}\right)^n.
\end{aligned}
\label{eq:exact-tensor-normsquared}
\end{equation}
Therefore
\begin{equation}
\upsilon_2(\mathcal M_{+,d}^{\otimes n})
= \bigl( \upsilon_2(\mathcal M_{+,d})\bigr)^n.
\label{eq:tensor-multipli}
\end{equation}
The value is attained by
$\Lambda_\star^{\otimes n}$, although the optimization allows an arbitrary joint CPTP map on all $n$ copies.
\end{theorem}

\begin{proof}
We already have the lower bound
\begin{equation}
\upsilon_2(\mathcal R_{+,d}^{\otimes n})^2\ge g_d^n,
\label{eq:tens-lower-bound-repeat}
\end{equation}
from the product choice
$\Lambda_\star^{\otimes n}$. Then it only remains to prove the opposite inequality for an arbitrary joint channel. Let
\begin{equation}
\Lambda^{(n)}:
\mathcal L((\mathsf H_+')^{\otimes n})
\longrightarrow \mathcal L(E)
\end{equation}
be any CPTP map, with $E$ finite dimensional, and let
$Q_{\Lambda^{(n)}}$ be the operator defined in
Equation \eqref{eq:joint-channel-kernel}. From Equation \eqref{eq:joint-kern_prop}, we have
\begin{equation}
Q_{\Lambda^{(n)}}\le I,
\qquad \mathscr T_n(Q_{\Lambda^{(n)}})\ge0,
\qquad \mathscr R_n(Q_{\Lambda^{(n)}})\ge0.
\end{equation}
By Lemma \ref{lem:tensor-comparison},
\begin{equation}
\mathsf L_d^{\otimes n} - K_d^{\otimes n} = \eta_d \left(\mathscr T_n(Z_{d,n}) + \mathscr R_n^\dagger(Z_{d,n}) \right),
\end{equation}
with
\begin{equation}
Z_{d,n}\ge0.
\end{equation}
Taking the trace against $Q_{\Lambda^{(n)}}$ gives
\begin{equation}
\begin{aligned}
& \operatorname{Tr} \left[ (\mathsf L_d^{\otimes n}-K_d^{\otimes n}) Q_{\Lambda^{(n)}} \right]\\
&=\eta_d\Bigl(\operatorname{Tr}[
Z_{d,n}\mathscr T_n(Q_{\Lambda^{(n)}})]+\operatorname{Tr}
[Z_{d,n}\mathscr R_n(Q_{\Lambda^{(n)}})]\Bigr)\\
&\ge0.
\end{aligned}
\end{equation}
Therefore
\begin{equation}
\operatorname{Tr} \left[K_d^{\otimes n}Q_{\Lambda^{(n)}}
\right] \le \operatorname{Tr} \left[
\mathsf L_d^{\otimes n}Q_{\Lambda^{(n)}}\right].
\end{equation}
Since $\mathsf L_d^{\otimes n}\ge0$ and $Q_{\Lambda^{(n)}}\le I,$ we also have
\begin{equation}
\operatorname{Tr}
\left[\mathsf L_d^{\otimes n}Q_{\Lambda^{(n)}}
\right] \le \operatorname{Tr}
\left(\mathsf L_d^{\otimes n}\right).
\end{equation}
The trace factorizes as
\begin{equation}
\operatorname{Tr} \left(\mathsf L_d^{\otimes n}
\right) =\bigl(\operatorname{Tr}\mathsf L_d\bigr)^n.
\end{equation}
Using Equation \eqref{ew:TR-L_d}, we obtain 
\begin{equation}
\operatorname{Tr} \left[
K_d^{\otimes n}Q_{\Lambda^{(n)}}\right]\le D_d^{2n}g_d^n.
\label{eq:tens-trace bound}
\end{equation}
The $n$ copy purity formula from
Equation \eqref{eq:tensor-kern objec} now gives
\begin{equation}
\begin{aligned}
 \left\|(\mathcal R_{+,d}^{\otimes n}\otimes\Lambda^{(n)})
(\Phi_{D_d}^{\otimes n})
\right\|_2^2 &=\frac1{D_d^{2n}} \operatorname{Tr}\left[
K_d^{\otimes n}Q_{\Lambda^{(n)}} \right]\\
&\le g_d^n.
\end{aligned}
\end{equation}
Since this holds for every finite dimensional $E$ and every joint CPTP map $\Lambda^{(n)}$,
\begin{equation}
\upsilon_2(\mathcal R_{+,d}^{\otimes n})^2
\le g_d^n.
\label{eq:tens-upperbound}
\end{equation}
Together with the lower bound,
\begin{equation}
\upsilon_2(\mathcal R_{+,d}^{\otimes n})^2= g_d^n.
\end{equation}
Finally, the exact symmetric reduction
Equation \eqref{eq:norm-suppo_reduc} gives
\begin{equation}
\upsilon_2(\mathcal M_{+,d}^{\otimes n})^2= g_d^n.
\end{equation}
Taking square roots and using Theorem \ref{thm:sw-one copy-norm} gives
\begin{equation}
\upsilon_2(\mathcal M_{+,d}^{\otimes n})
= \bigl( \upsilon_2(\mathcal M_{+,d})\bigr)^n.
\end{equation}
\end{proof}

\section{R\'enyi-2 entanglement of purification and additivity} \label{sec:7:eop.additiv}
We now return to the original purification problem. For a state $\rho$, the R\'enyi-$2$ entropy is
\begin{equation}
S_2(\rho):=-\log_2\operatorname{Tr}(\rho^2).
\label{eq:renyi-two-entropy}
\end{equation}
The R\'enyi-$2$ entanglement of purification of a bipartite state $\rho_{AC}$ is
\begin{equation}
\mathrm{EoP}_2(A:C)_\rho := \inf_{\substack{
|\varphi\rangle_{ACEE'}\\
\operatorname{Tr}_{EE'}
|\varphi\rangle\langle\varphi|=\rho_{AC}}} S_2(\varphi_{AE}),
\label{eq:renyi-eop-def}
\end{equation}
where
\begin{equation}
\varphi_{AE} :=\operatorname{Tr}_{CE'}
|\varphi\rangle\langle\varphi|.
\end{equation}
The optimization allows arbitrary finite dimensional auxiliary systems $E$ and $E'$. 

To connect this problem with the constrained norm studied above, we first choose one fixed purification of $W_{+,d}$. Since $W_{+,d}$ is maximally mixed on the $D_d$ dimensional symmetric subspace, a natural choice is
\begin{equation}
|\chi_d\rangle_{AC\mathsf H_+'} :=\frac1{\sqrt{D_d}}
\sum_{a=1}^{D_d}|e_a\rangle_{AC} \otimes
|e_a\rangle_{\mathsf H_+'},
\label{eq:ref-pur}
\end{equation}
where $\{|e_a\rangle_{AC}\}_{a=1}^{D_d}$ is the copy on
$A\otimes C$ of the symmetric basis introduced in
Equation \eqref{eq:sym-basis}. Tracing out the reference system gives
\begin{equation}
\begin{aligned}
\operatorname{Tr}_{\mathsf H_+'} |\chi_d\rangle\langle\chi_d| &= \frac1{D_d}
\sum_{a=1}^{D_d} |e_a\rangle\langle e_a|_{AC}\\
&=\frac{P_+^{AC}}{D_d}=W_{+,d}.
\end{aligned}
\end{equation}
Then $ |\chi_d\rangle $ is indeed a purification of $W_{+,d}$.
The other reduction of this purification is the operator that appeared in the constrained norm problem. And we have
\begin{equation}
\begin{aligned}
\operatorname{Tr}_C |\chi_d\rangle\langle\chi_d|
&= \frac1{D_d} \sum_{a,b=1}^{D_d} \operatorname{Tr}_C
\left( |e_a\rangle\langle e_b|_{AC} \right) \otimes
|e_a\rangle\langle e_b|_{\mathsf H_+'}.
\end{aligned}
\end{equation}
Under the identification of $A\otimes C$ with
$B_1\otimes B_2$,
\begin{equation}
\operatorname{Tr}_C \left( |e_a\rangle\langle e_b|_{AC}
\right) = \mathcal R_{+,d}
\left( |e_a\rangle\langle e_b|_{\mathsf H_+}\right).
\end{equation}
Therefore
\begin{equation}
\operatorname{Tr}_C|\chi_d\rangle\langle\chi_d|
=\Theta_{\mathcal R_{+,d}}.
\label{eq:pur-choi-margin}
\end{equation}
This is the link we need: before changing the purification, the $AE$ type marginal is exactly the normalized Choi operator of
$\mathcal R_{+,d}$. We next show that allowing all purifications is equivalent to allowing all CPTP maps on the reference system.
\begin{lemma}[Purifications as reference-system channels]
\label{lem:eop-norm-relation}

For every integer $n\ge1$,
\begin{equation}
\mathrm{EoP}_2 \left(A^{\otimes n}:C^{\otimes n}
\right)_{W_{+,d}^{\otimes n}}=-\log_2\left[
\upsilon_2\left(\mathcal R_{+,d}^{\otimes n}\right)^2
\right].
\label{eq:eop-r-norm-relation}
\end{equation}
Consequently,
\begin{equation}
\mathrm{EoP}_2\left(A^{\otimes n}:C^{\otimes n}
\right)_{W_{+,d}^{\otimes n}}=-\log_2\left[
\upsilon_2 \left(\mathcal M_{+,d}^{\otimes n}
\right)^2\right].
\label{eq:eop-norm-relation}
\end{equation}
\end{lemma}
\begin{proof}
The state $|\chi_d\rangle^{\otimes n}$ is a purification of $W_{+,d}^{\otimes n}$, with purifying system
$(\mathsf H_+')^{\otimes n}$. Now let
\begin{equation}
|\varphi\rangle_{A^{\otimes n}C^{\otimes n}EE'}
\end{equation}
be any other purification of $W_{+,d}^{\otimes n}$.

Two purifications of the same state are related by an isometry on the purifying system. Therefore there is an isometry
\begin{equation}
U:(\mathsf H_+')^{\otimes n} \longrightarrow E\otimes E'
\end{equation}
such that
\begin{equation}
|\varphi\rangle = \left(
I_{A^{\otimes n}C^{\otimes n}}
\otimes U \right) |\chi_d\rangle^{\otimes n}.
\label{eq:pur_from-ref}
\end{equation}
This is the usual isometric equivalence of purifications; (see, for example,
\cite{Watrous2018TheTheoryofQuantumInformation,Wilde2017QuantumInformationTheory}). Since the entropy in Equation \eqref{eq:renyi-eop-def} depends only on the subsystem $A^{\otimes n}E$, we trace out $E'$ and we define
\begin{equation}
\Lambda^{(n)}(Z)
:= \operatorname{Tr}_{E'} \left(UZU^\dagger\right).
\label{eq:purification-channel}
\end{equation}
Because $U$ is an isometry, $\Lambda^{(n)}$ is CPTP. By Equation \eqref{eq:pur-choi-margin}, the reduced state associated with $ |\varphi\rangle $ is therefore
\begin{equation}
\varphi_{A^{\otimes n}E} = \left( \operatorname{id}_{A^{\otimes n}} \otimes \Lambda^{(n)}
\right) \left( \Theta_{\mathcal R_{+,d}}^{\otimes n}
\right).
\label{eq:purific-out_chan}
\end{equation}
Thus every purification gives a CPTP map
$\Lambda^{(n)}$ of exactly the type appearing in the constrained norm optimization.

We also need the converse, let $\Lambda^{(n)}:
\mathcal L((\mathsf H_+')^{\otimes n})
\longrightarrow \mathcal L(E)$ be any CPTP map. By the Stinespring representation, there exist a finite dimensional system $E'$ and an isometry $U:(\mathsf H_+')^{\otimes n} \longrightarrow
E\otimes E'$ such that
\begin{equation}
\Lambda^{(n)}(Z) = \operatorname{Tr}_{E'}
\left( UZU^\dagger\right).
\end{equation}
Applying $U$ to the reference purification therefore produces a purification of $W_{+,d}^{\otimes n}$ whose
$A^{\otimes n}E$ reduction is precisely
Equation \eqref{eq:purific-out_chan}. Therefore the reduced states appearing in the entanglement of purification optimization are exactly the states
\begin{equation}
\left( \operatorname{id}_{A^{\otimes n}} \otimes
\Lambda^{(n)} \right) \left( \Theta_{\mathcal R_{+,d}}^{\otimes n} \right),
\end{equation}
where $E$ is arbitrary and $\Lambda^{(n)}$ is an arbitrary joint CPTP map. Therefore
\begin{equation}
\begin{aligned}
& \mathrm{EoP}_2\left(A^{\otimes n}:C^{\otimes n}
\right)_{W_{+,d}^{\otimes n}}= -\log_2 \left[\sup_{\Lambda^{(n)}}\left\|\left(\operatorname{id}_{A^{\otimes n}}\otimes\Lambda^{(n)}
\right)\left(\Theta_{\mathcal R_{+,d}}^{\otimes n}\right)
\right\|_2^2 \right].
\end{aligned}
\end{equation}
By the channel formulation of the constrained norm,
\begin{equation}
\sup_{\Lambda^{(n)}} \left\| \left(
\operatorname{id}_{A^{\otimes n}} \otimes
\Lambda^{(n)} \right) \left(\Theta_{\mathcal R_{+,d}}^{\otimes n} \right)\right\|_2^2=
\upsilon_2\left(\mathcal R_{+,d}^{\otimes n}\right)^2.
\end{equation}
This proves Equation \eqref{eq:eop-r-norm-relation}. The exact symmetric reduction then gives
\begin{equation}
\upsilon_2 \left(
\mathcal M_{+,d}^{\otimes n} \right)
= \upsilon_2\left(\mathcal R_{+,d}^{\otimes n}\right),
\end{equation}
which proves Equation \eqref{eq:eop-norm-relation}.
\end{proof}
We can now insert the tensor power norm computed in the previous subsection.

\begin{theorem}[Exact value and tensor power additivity]
\label{thm:eop-additi}
For every integer $d\ge2$,
\begin{equation}
\mathrm{EoP}_2(A:C)_{W_{+,d}} = -\log_2 g_d
= \log_2 \left( \frac{4d(d+1)}{d^2+3d+4}\right).
\label{eq:eop-one-copy}
\end{equation}
Moreover, for every integer $n\ge1$,
\begin{equation}
\mathrm{EoP}_2
\left( A^{\otimes n}:C^{\otimes n}
\right)_{W_{+,d}^{\otimes n}}
= n\, \mathrm{EoP}_2(A:C)_{W_{+,d}}.
\label{eq:eop-tensor-add}
\end{equation}
The minimum is attained by a product of one-copy purifications. For $n$ copies, one may take
\begin{equation}
\dim E=\dim E'=d^n.
\end{equation}
\end{theorem}

\begin{proof}
By Lemma~\ref{lem:eop-norm-relation} and
Theorem~\ref{thm:tensor-multipli},
\begin{align}
& \mathrm{EoP}_2 \left(A^{\otimes n}:C^{\otimes n}\right)_{W_{+,d}^{\otimes n}} =-\log_2\left[ \upsilon_2 \left( \mathcal R_{+,d}^{\otimes n} \right)^2 \right]\nonumber \\
&= -\log_2(g_d^n) = -n\log_2 g_d.
\end{align}
For $n=1$, we obtain
\begin{equation}
\mathrm{EoP}_2(A:C)_{W_{+,d}}
=\log_2\left(\frac{4d(d+1)}{d^2+3d+4}\right).
\end{equation}
The formula for general $n$ then immediately gives
\begin{equation}
\mathrm{EoP}_2\left(A^{\otimes n}:C^{\otimes n}
\right)_{W_{+,d}^{\otimes n}}= n\, \mathrm{EoP}_2(A:C)_{W_{+,d}}.
\end{equation}
It remains only to identify a purification that attains this value. For the one copy optimizing channel \eqref{eq:lower-postpr} isometric dilation is simply the inclusion
\begin{equation}
J_+': \mathsf H_+' \longrightarrow B_1'\otimes B_2'.
\end{equation}
Applying this isometry to the reference purification gives
\begin{equation}
|\varphi_{\star,d}\rangle_{ACB_1'B_2'}
:=\left(I_{AC}\otimes J_+'\right)
|\chi_d\rangle_{AC\mathsf H_+'}.
\label{eq:optimal-purifi}
\end{equation}
Tracing out $C$ and $B_2'$ gives
\begin{equation}
\operatorname{Tr}_{CB_2'} |\varphi_{\star,d}\rangle
\langle\varphi_{\star,d}| = \omega_\star.
\end{equation}
Since $\operatorname{Tr}(\omega_\star^2)=g_d,$ this purification attains the one copy minimum.

For $n$ copies, let take $|\varphi_{\star,d}\rangle^{\otimes n}$. Its reduced state on $A^{\otimes n}\otimes(B_1')^{\otimes n}$ is $\omega_\star^{\otimes n}$ whose purity is
\begin{equation}
\operatorname{Tr}\left[(\omega_\star^{\otimes n})^2
\right]=g_d^n.
\end{equation}
Thus, the product purification attains the exact $n$ copy value, even though the optimization allows arbitrary joint purifications across all $n$ copies. Here we may take
\begin{equation}
E=(B_1')^{\otimes n},
\qquad E'=(B_2')^{\otimes n},
\end{equation}
so $\dim E=\dim E'=d^n.$
\end{proof}

The tensor power theorem treated several copies with the same local dimension. The same argument also applies when the dimensions are different. The only change is that the one copy relation for the $k$-th factor contains its own coefficient $\eta_{d_k}$, so this coefficient must remain attached to that factor in the product construction.

\begin{corollary}[Additivity across different local dimensions]
\label{cor:sw-mixed_dim.add}

Let $r\ge1$ and let $d_1,\ldots,d_r\ge2$. For each $k$, let
\begin{equation}
A_k\simeq C_k\simeq\mathbb C^{d_k},
\end{equation}
and let $W_{+,d_k}$ be the normalized symmetric Werner state on $A_k\otimes C_k$. Define
\begin{equation}
\rho :=\bigotimes_{k=1}^r W_{+,d_k},
\end{equation}
with bipartition
\begin{equation}
A_1\cdots A_r:C_1\cdots C_r.
\end{equation}
Then
\begin{equation}
\begin{aligned}
\mathrm{EoP}_2
(A_1\cdots A_r:C_1\cdots C_r)_\rho&=
\sum_{k=1}^r \mathrm{EoP}_2(A_k:C_k)_{W_{+,d_k}}\\
&= \sum_{k=1}^r \log_2 \left( \frac{4d_k(d_k+1)}
{d_k^2+3d_k+4} \right).
\end{aligned}
\label{eq:mixed-dim.add}
\end{equation}
The product $\bigotimes_{k=1}^r |\varphi_{\star,d_k}\rangle$ is optimal even when arbitrary joint finite dimensional purifications of $\rho$ are allowed.
\end{corollary}

\begin{proof}
For each $k$, let
\begin{equation}
D_{d_k}=\frac{d_k(d_k+1)}2,
\end{equation}
and define
\begin{equation}
\mathcal N := \bigotimes_{k=1}^r
\mathcal R_{+,d_k}, \qquad D:=\prod_{k=1}^rD_{d_k}.
\end{equation}
The product of the reference purifications from
Equation \eqref{eq:ref-pur}, and
$\bigotimes_{k=1}^r|\chi_{d_k}\rangle,$ is a purification of $\rho$. After tracing out $C_1\cdots C_r$, its remaining state is
\begin{equation}
\bigotimes_{k=1}^r \Theta_{\mathcal R_{+,d_k}} =
\Theta_{\mathcal N}.
\end{equation}
Therefore the same purification to channel argument used in
Lemma \ref{lem:eop-norm-relation} gives
\begin{equation}
\mathrm{EoP}_2 (A_1\cdots A_r:C_1\cdots C_r)_\rho
= -\log_2 \left[ \upsilon_2(\mathcal N)^2\right].
\label{eq:mixed-pur norm}
\end{equation}
Now we determine $\upsilon_2(\mathcal N)^2$.

The lower bound is immediate,  for the $k$-th factor, choose the one copy optimizing channel
$\Lambda_{\star,d_k}$. Their product produces
\begin{equation}
\bigotimes_{k=1}^r\omega_{\star,d_k},
\end{equation}
whose purity is $\prod_{k=1}^r g_{d_k}.$ Therefore
\begin{equation}
\upsilon_2(\mathcal N)^2 \ge \prod_{k=1}^r g_{d_k}.
\label{eq:mixed-lower-bound}
\end{equation}
We now prove that no joint post-processing channel can do better. Let
\begin{equation}
\Lambda: \mathcal L \left(
\bigotimes_{k=1}^r\mathsf H_{+,d_k}' \right)
\longrightarrow \mathcal L(E)
\end{equation}
be an arbitrary CPTP map. No product form is assumed. Define its channel kernel as 
\begin{equation}
Q_\Lambda=(\Lambda^\dagger\otimes\Lambda^\dagger)
(\Pi_{E,\widetilde E}).
\end{equation}
To write the partial transpose and realignment factor by factor, group each symmetric input with its copy and let
$\mathscr T_k$ and $\mathscr R_k$ denote the transformations for the
$d_k$ dimensional factor. Define
\begin{equation}
\mathscr T_{\boldsymbol d}
:= \bigotimes_{k=1}^r\mathscr T_k, \qquad
\mathscr R_{\boldsymbol d} := \bigotimes_{k=1}^r\mathscr R_k,
\label{eq:mixed-transforms}
\end{equation}
where $\boldsymbol d=(d_1,\ldots,d_r).$ Treating the full product reference space as one input system,
Lemma \ref{lem:1channel-k-props} gives
\begin{equation}
Q_\Lambda\le I, \qquad
\mathscr T_{\boldsymbol d}(Q_\Lambda)\ge0,\qquad
\mathscr R_{\boldsymbol d}(Q_\Lambda)\ge0.
\label{eq:mixed-channel-ker-prop}
\end{equation}
The fixed kernel of $\mathcal N$ is $\bigotimes_{k=1}^rK_{d_k}.$ Thus the purity associated with $\Lambda$ is
\begin{equation}
\frac1{D^2} \operatorname{Tr}
\left[ \left( \bigotimes_{k=1}^rK_{d_k}\right)
Q_\Lambda \right],
\label{eq:mixed-kernel-object}
\end{equation}
where we used $K_{d_k}^T=K_{d_k}$ for every $k$.

We therefore need to compare $\bigotimes_{k=1}^rK_{d_k}$
with $\bigotimes_{k=1}^r\mathsf L_{d_k}.$ As in the tensor power proof, adding and subtracting the intermediate products gives

\begin{equation}
\begin{aligned}
&\bigotimes_{k=1}^r\mathsf L_{d_k}-\bigotimes_{k=1}^rK_{d_k} =
\sum_{k=1}^r\left(
\bigotimes_{j<k}\mathsf L_{d_j}\right)\otimes
(\mathsf L_{d_k}-K_{d_k})\otimes
\left(\bigotimes_{j>k}K_{d_j}\right).
\end{aligned}
\label{eq:mixed-telescop}
\end{equation}
For the $k$-th middle factor, the one copy relation is
\begin{equation}
\mathsf L_{d_k}-K_{d_k}
= \eta_{d_k} \left( \mathscr T_k(Z_{d_k}) +
\mathscr R_k^\dagger(Z_{d_k}) \right).
\end{equation}
The coefficient $\eta_{d_k}$ now depends on $k$. This is why it must be included inside the corresponding term.
Then let define
\begin{equation}
\begin{aligned}
Z_{\boldsymbol d}:= \sum_{k=1}^r
& \left( \bigotimes_{j<k}
\mathscr T_j(\mathsf L_{d_j}) \right) \otimes
\bigl(\eta_{d_k}Z_{d_k}\bigr) \otimes\left(
\bigotimes_{j>k}\mathscr T_j(K_{d_j})\right),
\end{aligned}
\label{eq:mixed-positive_op}
\end{equation}
where empty tensor products are omitted. By Lemma \ref{lem:tensor.positive.factor},
\begin{equation}
\mathscr T_j(\mathsf L_{d_j})\ge0,
\qquad \mathscr T_j(K_{d_j})\ge0,
\end{equation}
and by Lemma \ref{lem:positive-Z},
\begin{equation}
Z_{d_k}\ge0. \qquad \text { and } \qquad \eta_{d_k}>0.
\end{equation}
Therefore every term in Equation \eqref{eq:mixed-positive_op} is positive, and therefore
\begin{equation}
Z_{\boldsymbol d}\ge0.
\label{eq:mixed-Z-positive}
\end{equation}
We now apply the two transformations. From Equation \eqref{eq:common-preimages},
\begin{equation}
\mathscr T_j \bigl( \mathscr T_j(\mathsf L_{d_j})
\bigr) = \mathscr R_j^\dagger \bigl(
\mathscr T_j(\mathsf L_{d_j})\bigr)=\mathsf L_{d_j},
\end{equation}
and similarly,
\begin{equation}
\mathscr T_j \bigl( \mathscr T_j(K_{d_j})\bigr)
=\mathscr R_j^\dagger\bigl(\mathscr T_j(K_{d_j})\bigr)
=K_{d_j}.
\end{equation}
Therefore
\begin{equation}
\begin{aligned}
&\mathscr T_{\boldsymbol d}(Z_{\boldsymbol d})
+\mathscr R_{\boldsymbol d}^\dagger(Z_{\boldsymbol d})=\\
&\sum_{k=1}^r\left(\bigotimes_{j<k}\mathsf L_{d_j}\right) \quad\otimes\eta_{d_k}
\left(\mathscr T_k(Z_{d_k})+ \mathscr R_k^\dagger(Z_{d_k})
\right) \quad\otimes
\left(\bigotimes_{j>k}K_{d_j}\right).
\end{aligned}
\end{equation}
Using the one copy relation in the middle factor gives
\begin{equation}
\begin{aligned}
\mathscr T_{\boldsymbol d}(Z_{\boldsymbol d})
+ \mathscr R_{\boldsymbol d}^\dagger(Z_{\boldsymbol d})
= \sum_{k=1}^r \left(\bigotimes_{j<k}\mathsf L_{d_j} \right)\otimes
(\mathsf L_{d_k}-K_{d_k})\otimes\left(
\bigotimes_{j>k}K_{d_j}\right).
\end{aligned}
\end{equation}
By Equation \eqref{eq:mixed-telescop},
\begin{equation}
\mathscr T_{\boldsymbol d}(Z_{\boldsymbol d})
+\mathscr R_{\boldsymbol d}^\dagger(Z_{\boldsymbol d})
=\bigotimes_{k=1}^r\mathsf L_{d_k}
- \bigotimes_{k=1}^rK_{d_k}.
\label{eq:mixed-comparison}
\end{equation}
We can now pair this identity with $Q_\Lambda$. Using Equations \eqref{eq:mixed-channel-ker-prop} and
\eqref{eq:mixed-Z-positive},
\begin{equation}
\begin{aligned}
&\operatorname{Tr}
\left[\left(\bigotimes_{k=1}^r\mathsf L_{d_k}
-\bigotimes_{k=1}^rK_{d_k}
\right)Q_\Lambda\right]\\
&=\operatorname{Tr}
\left[Z_{\boldsymbol d}
\mathscr T_{\boldsymbol d}(Q_\Lambda)\right]
+ \operatorname{Tr}
\left[ Z_{\boldsymbol d}
\mathscr R_{\boldsymbol d}(Q_\Lambda)\right] \ge0.
\end{aligned}
\end{equation}
Therefore
\begin{equation}
\operatorname{Tr}
\left[\left(\bigotimes_{k=1}^rK_{d_k}\right)Q_\Lambda \right]
\le \operatorname{Tr}\left[
\left(\bigotimes_{k=1}^r\mathsf L_{d_k}
\right)Q_\Lambda\right].
\end{equation}
Since $\bigotimes_{k=1}^r\mathsf L_{d_k}\ge0$ and $Q_\Lambda\le I$,
\begin{equation}
\begin{aligned}
\operatorname{Tr}
\left[\left(\bigotimes_{k=1}^r\mathsf L_{d_k}\right)
Q_\Lambda\right]
&\le\operatorname{Tr}\left(
\bigotimes_{k=1}^r\mathsf L_{d_k}\right) = \prod_{k=1}^r
\operatorname{Tr}\mathsf L_{d_k}.
\end{aligned}
\end{equation}
Considering $\operatorname{Tr}\mathsf L_{d_k}=D_{d_k}^2g_{d_k},$
we get
\begin{equation}
\prod_{k=1}^r\operatorname{Tr}\mathsf L_{d_k}=D^2
\prod_{k=1}^r g_{d_k}.
\end{equation}
Substituting this into Equation \eqref{eq:mixed-kernel-object} gives
\begin{equation}
\upsilon_2(\mathcal N)^2 \le \prod_{k=1}^r g_{d_k}.
\end{equation}
Together with
Equation \eqref{eq:mixed-lower-bound},
\begin{equation}
\upsilon_2(\mathcal N)^2=\prod_{k=1}^r g_{d_k}.
\label{eq:mixed-exact-norm}
\end{equation}
Finally, Equation \eqref{eq:mixed-pur norm} gives
\begin{equation}
\begin{aligned}
&\mathrm{EoP}_2
(A_1\cdots A_r:C_1\cdots C_r)_\rho\\
&= -\log_2 \left( \prod_{k=1}^r g_{d_k}
\right)= \sum_{k=1}^r\left(-\log_2g_{d_k}\right)=\sum_{k=1}^r
\mathrm{EoP}_2(A_k:C_k)_{W_{+,d_k}}.
\end{aligned}
\end{equation}
Using the one copy formula from
Theorem \ref{thm:eop-additi} gives
Equation \eqref{eq:mixed-dim.add}.
The lower bound was attained by the product channels
$\bigotimes_k\Lambda_{\star,d_k}$. Their isometric dilations are the product of the inclusions $J_{+,d_k}'$, so the corresponding purification is
\begin{equation}
\bigotimes_{k=1}^r |\varphi_{\star,d_k}\rangle.
\end{equation}
Therefore this product purification attains the minimum even though arbitrary joint purifications are allowed.
\end{proof}
Thus, the optimal product construction extends to arbitrary finite products of normalized symmetric Werner states, including factors with different local dimensions. This corollary does not imply uniqueness of the optimizer, minimality of the auxiliary dimensions, additivity with states outside this family, or a corresponding result for other Werner parameters.

\section{Stability near the symmetric endpoint}
\label{sec:8:stability}
The exact results above concern the normalized symmetric Werner state, which is supported entirely on the symmetric subspace. We now ask what happens when a small amount of weight is moved into the antisymmetric subspace. Here we address whether the optimal value $\mathrm{EoP}_2$ remains close to its value at the symmetric endpoint. We also show that the canonical purification continues to achieve an entropy close to the optimum, but we do not claim that every optimizer must be close to it.

Let fix $d\ge2$ and define the antisymmetric projection and the normalized antisymmetric Werner state as
\begin{equation}
P_-^{AC}:=\frac{I_{AC}-\Pi_{AC}}2,
\qquad D_{-,d} := \frac{d(d-1)}2, \qquad
W_{-,d} := \frac{P_-^{AC}}{D_{-,d}}.
\label{eq:antisym-state}
\end{equation}
For $0\le\varepsilon\le1$, consider
\begin{equation}
\rho_\varepsilon:= (1-\varepsilon)W_{+,d}+ \varepsilon W_{-,d}.
\label{eq:pert_state}
\end{equation}
Thus $\varepsilon$ is exactly the weight placed in the antisymmetric subspace. Since $W_{+,d}$ and $W_{-,d}$ have orthogonal supports,
\begin{equation}
\frac12 \left\|
\rho_\varepsilon-W_{+,d}
\right\|_1= \varepsilon.
\label{eq:perturb-distance}
\end{equation}
In the Werner parametrization
\begin{equation}
W_d(t)=\frac{1-t}{d^2}I_{AC}+\frac td\Pi_{AC},
\end{equation}
the state $\rho_\varepsilon$ corresponds to
\begin{equation}
t=\frac1{d+1}-\frac{2d}{d^2-1}\varepsilon.
\label{eq:perturb-parameter}
\end{equation}
It will be more convenient to compare optimal purities before taking logarithms. Then we define
\begin{equation}
q_d(\varepsilon):=2^{-\mathrm{EoP}_2(A:C)_{\rho_\varepsilon}}.
\label{eq:pert_optimal-purity}
\end{equation}
By the definition of $\mathrm{EoP}_2$, $q_d(\varepsilon)$ is the supremum of $\operatorname{Tr}(\varphi_{AE}^2)$ over all finite dimensional purifications of $\rho_\varepsilon$. It is therefore an optimal reduced purity, not the purity of $\rho_\varepsilon$ itself. At the symmetric endpoint,
\begin{equation}
q_d(0)=g_d.
\label{eq:endpoint-opt-pur}
\end{equation}
We first show that this optimal value cannot change discontinuously when $\varepsilon$ is small.

\begin{proposition}[Stability of the optimal value]
\label{prop:sw-value-stability}
For every integer $d\ge2$ and every
$0\le\varepsilon\le1$,
\begin{equation}
|q_d(\varepsilon)-g_d|\le 2\sqrt{\varepsilon}.
\label{eq:p-stab}
\end{equation}
In particular,
\begin{equation}
q_d(\varepsilon)\longrightarrow g_d
\end{equation}
and
\begin{equation}
\mathrm{EoP}_2(A:C)_{\rho_\varepsilon}
\longrightarrow \mathrm{EoP}_2(A:C)_{W_{+,d}}
\end{equation}
as $\varepsilon\downarrow0$.
\end{proposition}
\begin{proof}
We want to compare the two purification optimizations using the same post-processing channel. For this purpose, use the common reference system
\begin{equation}
H'=B_1'\otimes B_2'\simeq \mathbb C^d\otimes\mathbb C^d.
\end{equation}
For each $\varepsilon$, define
\begin{equation}
|\varphi_\varepsilon\rangle
:= \sum_{i,j=1}^d \left(\sqrt{\rho_\varepsilon}\,
|i,j\rangle_{AC}\right) \otimes
|i,j\rangle_{B_1'B_2'}.
\label{eq:pert-can_pur}
\end{equation}
Its reduction to $A\otimes C$ is $\rho_\varepsilon$, so it is a purification of $\rho_\varepsilon$. At $\varepsilon=0$,
\begin{equation}
|\varphi_0\rangle=|\varphi_{\star,d}\rangle
\end{equation}
from Equation \eqref{eq:optimal-purifi}. For a CPTP map $\Lambda: \mathcal L(H')
\longrightarrow \mathcal L(E),$ we define
\begin{equation}
\omega_\varepsilon^\Lambda:=(\operatorname{id}_A\otimes\Lambda)
\left(\operatorname{Tr}_C|\varphi_\varepsilon\rangle
\langle\varphi_\varepsilon| \right).
\label{eq:pert_channel-output}
\end{equation}
As in the purification to channel argument above, every reduced state entering the purification optimization can be obtained in this way.
For $0<\varepsilon<1$, the reference state has full support on $H'$. At the rank deficient endpoints, the channel initially is needed only on the support of the reference state. Any such channel can be extended to all of $H'$ by sending the orthogonal complement to a fixed state. Therefore, we may formulate the optimization for every $\varepsilon$ over the same set of CPTP maps on $H'$. This allows us to compare $\omega_\varepsilon^\Lambda$ and
$\omega_0^\Lambda$ for the same $\Lambda$. Now because the symmetric and antisymmetric subspaces are orthogonal,
\begin{equation}
\sqrt{\rho_\varepsilon}=\sqrt{\frac{1-\varepsilon}{D_d}}\,P_+^{AC} +\sqrt{\frac{\varepsilon}{D_{-,d}}}\,P_-^{AC}.
\end{equation}
Since $\sqrt{W_{+,d}}=\frac{P_+^{AC}}{\sqrt{D_d}},$ we obtain
\begin{equation}
\begin{aligned}
\langle \varphi_0 |
\varphi_\varepsilon \rangle
&=\operatorname{Tr}\left(\sqrt{W_{+,d}}
\sqrt{\rho_\varepsilon}\right)=\sqrt{1-\varepsilon}.
\end{aligned}
\end{equation}
For two pure states, the trace distance is therefore
\begin{equation}
\left\| |\varphi_\varepsilon\rangle
\langle\varphi_\varepsilon|- |\varphi_0\rangle \langle\varphi_0| \right\|_1 = 2\sqrt{1-
|\langle\varphi_0|\varphi_\varepsilon\rangle|^2}
=2\sqrt{\varepsilon}.
\end{equation}
Partial trace and CPTP maps contract the trace norm
\cite{Watrous2018TheTheoryofQuantumInformation}. Therefore, for every $\Lambda$, we can write
\begin{equation}
\left\| \omega_\varepsilon^\Lambda - \omega_0^\Lambda \right\|_1 \le 2\sqrt{\varepsilon}.
\label{eq:output-stability}
\end{equation}
We now translate this trace distance bound into a purity bound. Let $\omega$ and $\sigma$ be density operators. Since $\operatorname{Tr}(\omega-\sigma)=0,$ we may write
\begin{equation}
\begin{aligned}
\operatorname{Tr}(\omega^2) - \operatorname{Tr}(\sigma^2)
&= \operatorname{Tr} \left[(\omega-\sigma)(\omega+\sigma)\right] \\
&= \operatorname{Tr} \left[(\omega-\sigma)(\omega+\sigma-I)\right].
\end{aligned}
\end{equation}
Moreover, $0\le\omega+\sigma\le2I,$ so
\begin{equation}
\|\omega+\sigma-I\|_\infty\le1.
\end{equation}
Therefore
\begin{equation}
\left| \operatorname{Tr}(\omega^2) -
\operatorname{Tr}(\sigma^2) \right| \le \|\omega-\sigma\|_1.
\label{eq:pur-trace-distance}
\end{equation}
Applying this to $\omega_\varepsilon^\Lambda$ and
$\omega_0^\Lambda$, and using Equation \eqref{eq:output-stability}, gives
\begin{equation}
\left| \operatorname{Tr} \left[ (\omega_\varepsilon^\Lambda)^2 \right]-\operatorname{Tr}
\left[(\omega_0^\Lambda)^2\right]\right|\le
2\sqrt{\varepsilon}.
\end{equation}
This holds for every channel in the common optimization set. Therefore the suprema of the two purities can differ by at most the same amount as
\begin{equation}
|q_d(\varepsilon)-q_d(0)|\le 2\sqrt{\varepsilon}.
\end{equation}
And $q_d(0)=g_d$ proves Equation \eqref{eq:p-stab}. Finally, $g_d>0$, and the function $-\log_2 x$ is continuous at $x=g_d$. Therefore convergence of the optimal purities implies convergence of the corresponding R\'enyi-$2$ entanglement of purification.
\end{proof}
The proposition shows that the optimal value cannot jump when a small antisymmetric component is added. It does not tell us whether the $\sqrt{\varepsilon}$ dependence is merely an artifact of the estimate. To see what actually happens near the endpoint, we evaluate the same canonical purification $|\varphi_\varepsilon\rangle$ explicitly.
Now we define its reduced state
\begin{equation}
\omega_{\mathrm{can},d}(\varepsilon)
:= \operatorname{Tr}_{CB_2'}
|\varphi_\varepsilon\rangle \langle\varphi_\varepsilon|,
\label{eq:pert_canonical-output}
\end{equation}
and write
\begin{equation}
p_{\mathrm{can},d}(\varepsilon) :=\operatorname{Tr}
\left[\omega_{\mathrm{can},d}(\varepsilon)^2\right].
\end{equation}
Since this state comes from one particular purification,
\begin{equation}
p_{\mathrm{can},d}(\varepsilon) \le q_d(\varepsilon).
\label{eq:pert_canonical-bound}
\end{equation}
We now compute $p_{\mathrm{can},d}(\varepsilon)$. From Equations \eqref{eq:antisym-state} and \eqref{eq:sym-state} we can write
\begin{equation}
\sqrt{\rho_\varepsilon}
= u_\varepsilon I_{AC} +v_\varepsilon\Pi_{AC},
\label{eq:pert-sq}
\end{equation}
where
\begin{equation}
u_\varepsilon = \frac12 \left( \sqrt{\frac{1-\varepsilon}{D_d}} + \sqrt{\frac{\varepsilon}{D_{-,d}}} \right),
\qquad v_\varepsilon = \frac12 \left(
\sqrt{\frac{1-\varepsilon}{D_d}} -
\sqrt{\frac{\varepsilon}{D_{-,d}}}\right).
\label{eq:pert_uv}
\end{equation}
Substituting Equation \eqref{eq:pert-sq} into
Equation \eqref{eq:pert-can_pur} gives
\begin{equation}
|\varphi_\varepsilon\rangle
= \sum_{i,j=1}^d \left( u_\varepsilon|i,j\rangle_{AC} + v_\varepsilon|j,i\rangle_{AC} \right) \otimes |i,j\rangle_{B_1'B_2'}.
\label{eq:pert-puri-ex}
\end{equation}
To trace out $C$ and $B_2'$, consider matrix elements on
$A\otimes B_1'$. From Equation \eqref{eq:pert-puri-ex},

\begin{equation}
\begin{aligned}
&\langle a,b|
\omega_{\mathrm{can},d}(\varepsilon)|a',b'\rangle = v_\varepsilon^2 \delta_{aa'}\delta_{bb'} + \left( d u_\varepsilon^2 + 2u_\varepsilon v_\varepsilon \right) \delta_{ab}\delta_{a'b'}.
\end{aligned}
\end{equation}
Since
\begin{equation}
\langle a,b| \Phi_d^{AB_1'} |a',b'\rangle=
\frac1d \delta_{ab}\delta_{a'b'},
\end{equation}
we obtain
\begin{align}
\omega_{\mathrm{can},d}(\varepsilon)
&=v_\varepsilon^2 I_{AB_1'}
+\left(d^2u_\varepsilon^2+2d u_\varepsilon v_\varepsilon \right)\Phi_d^{AB_1'}\nonumber\\
&=v_\varepsilon^2\left(I_{AB_1'}-\Phi_d^{AB_1'} \right) + (du_\varepsilon+v_\varepsilon)^2 \Phi_d^{AB_1'}.
\label{eq:pert_canonical-output-explicit}
\end{align}
Thus, from the second term we know that the eigenvalue on the maximally entangled vector is $\lambda_d(\varepsilon)=(du_\varepsilon+v_\varepsilon)^2.$
Substituting Equation \eqref{eq:pert_uv} and simplifying gives
\begin{equation}
\lambda_d(\varepsilon)= \frac{ d+1-2\varepsilon+2\sqrt{(d^2-1)\varepsilon(1-\varepsilon)}}{2d}.
\label{eq:pert-cano eigen}
\end{equation}
The remaining $d^2-1$ eigenvalues are all equal to
$v_\varepsilon^2$. Since the state has trace one,
\begin{equation}
v_\varepsilon^2=\frac{1-\lambda_d(\varepsilon)}{d^2-1}.
\end{equation}
Therefore
\begin{equation}
p_{\mathrm{can},d}(\varepsilon)=\lambda_d(\varepsilon)^2
+\frac{\left(1-\lambda_d(\varepsilon)\right)^2}{d^2-1}.
\label{eq:pert_canonical-purity}
\end{equation}
This explicit achievable purity determines the order of the change at the endpoint.

\begin{corollary}[Square root change at the endpoint]
\label{cor:sq-stab}

For each fixed $d\ge2$, define
\begin{equation}
c_d :=\frac{(d+2)\sqrt{d^2-1}}{d(d+1)}>0.
\label{eq:sq-coef}
\end{equation}
Then, for all sufficiently small $\varepsilon>0$,
\begin{equation}
\frac{c_d}{2}\sqrt{\varepsilon}
\le q_d(\varepsilon)-g_d \le 2\sqrt{\varepsilon}.
\label{eq:two-side_sq}
\end{equation}
In particular, the optimal purity is continuous at the symmetric endpoint, but there is no finite constant $C_d$ for which
\begin{equation}
|q_d(\varepsilon)-g_d| \le C_d\varepsilon
\end{equation}
holds for all sufficiently small positive $\varepsilon$.
\end{corollary}

\begin{proof}
We first expand the eigenvalue
Equation \eqref{eq:pert-cano eigen}. As
$\varepsilon\downarrow0$,
\begin{equation}
\lambda_d(\varepsilon)
=\frac{d+1}{2d} + \frac{\sqrt{d^2-1}}{d} \sqrt{\varepsilon}
+ O(\varepsilon).
\label{eq:can-eigen}
\end{equation}
Now define
\begin{equation}
f_d(\lambda) := \lambda^2+ \frac{(1-\lambda)^2}{d^2-1},
\end{equation}
so that
\begin{equation}
p_{\mathrm{can},d}(\varepsilon)= f_d(\lambda_d(\varepsilon)).
\end{equation}
At $\lambda_d(0)= \frac{d+1}{2d},$ and we have $f_d(\lambda_d(0)) = g_d,$ and
\begin{equation}
\begin{aligned}
f_d'(\lambda_d(0))
&=2\lambda_d(0)- \frac{2(1-\lambda_d(0))}{d^2-1}= \frac{d+2}{d+1}.
\end{aligned}
\end{equation}
Combining this with
Equation \eqref{eq:can-eigen} gives
\begin{equation}
p_{\mathrm{can},d}(\varepsilon)
= g_d + c_d\sqrt{\varepsilon} + O(\varepsilon),
\label{eq:canonical-purity-expansion}
\end{equation}
with $c_d$ as in Equation \eqref{eq:sq-coef}. Since $c_d>0$, for all sufficiently small positive $\varepsilon$, we can write
\begin{equation}
p_{\mathrm{can},d}(\varepsilon)-g_d
\ge \frac{c_d}{2}\sqrt{\varepsilon}.
\end{equation}
Also, by Equuation \eqref{eq:pert_canonical-bound} we obtain
\begin{equation}
q_d(\varepsilon)-g_d
\ge \frac{c_d}{2}\sqrt{\varepsilon}.
\end{equation}
Now the upper bound follows from Proposition \ref{prop:sw-value-stability}
\begin{equation}
q_d(\varepsilon)-g_d \le |q_d(\varepsilon)-g_d| \le 2\sqrt{\varepsilon}.
\end{equation}
This proves Equation \eqref{eq:two-side_sq}. Finally, as $\varepsilon\downarrow0$
\begin{equation}
\frac{ q_d(\varepsilon)-g_d }{\varepsilon}\ge
\frac{c_d}{2\sqrt{\varepsilon}} \longrightarrow \infty
\end{equation}
Therefore no finite linear bound in
$\varepsilon$ can hold near the endpoint.
\end{proof}
We therefore have the sharper statement
\begin{equation}
q_d(\varepsilon)-g_d = \Theta_d(\sqrt{\varepsilon})
\qquad (\varepsilon\downarrow0),
\end{equation}
and since $-\log_2x$ is differentiable at $g_d>0$, the corresponding change in the R\'enyi-$2$ entanglement of purification is also of square-root order. In particular,
\begin{equation}
\mathrm{EoP}_2(A:C)_{W_{+,d}} - \mathrm{EoP}_2(A:C)_{\rho_\varepsilon}
= \Theta_d(\sqrt{\varepsilon}).
\end{equation}
Note that the coefficient $c_d$ above comes from the canonical purification. We have not determined the exact leading coefficient of the optimized value $q_d(\varepsilon)$.

The preceding estimates also quantify how close the canonical purification is to optimal in achieved entropy. From Proposition \ref{prop:sw-value-stability},
\begin{equation}
q_d(\varepsilon) \le \min \left\{ 1, g_d+2\sqrt{\varepsilon} \right\},
\end{equation}
while Equation \eqref{eq:pert_canonical-bound} gives $p_{\mathrm{can},d}(\varepsilon)
\le q_d(\varepsilon).$ And since we know $\mathrm{EoP}_2(A:C)_{\rho_\varepsilon} = -\log_2q_d(\varepsilon),$ we obtain
\begin{align}
&-\log_2 \left( \min \left\{ 1, g_d+2\sqrt{\varepsilon}
\right\} \right)\le \mathrm{EoP}_2(A:C)_{\rho_\varepsilon}
\le -\log_2 p_{\mathrm{can},d}(\varepsilon).
\label{eq:ent-stab-bound}
\end{align}
Both sides converge to
\begin{equation}
-\log_2g_d= \mathrm{EoP}_2(A:C)_{W_{+,d}}.
\end{equation}
Therefore the entropy achieved by the canonical purification and the minimum entropy differ by an amount that tends to zero as
$\varepsilon\downarrow0$.

These results concern the one copy optimal value along the Werner family near the symmetric endpoint. They do not establish exact optimality of the canonical purification for $\varepsilon>0$, additivity away from the endpoint, or nearly optimal purification to the canonical one.


\section{Summary and Conclusion}\label{sec:9:sum-con}

The central difficulty in entanglement of purification problems is not constructing a good purification, but establishing that no more general purification can do any better. At the symmetric Werner endpoint, we resolve this difficulty exactly for Rényi order two, and we showed for every $n\ge1$,
\begin{equation*}
\mathrm{EoP}_2
\left(A^{\otimes n}:C^{\otimes n}
\right)_{W_{+,d}^{\otimes n}}=n\,
\mathrm{EoP}_2(A:C)_{W_{+,d}}.    
\end{equation*}
The proof proceeds by converting the purification problem into a constrained output Schatten-$2$ norm, reducing the relevant map to the symmetric input space, and then solving the resulting optimization exactly.  The upper bound is obtained without assuming that the optimizing postprocessing channel is covariant, product, or otherwise restricted. The operator inequalities lead to the one copy bound and extends to arbitrary joint postprocessing on tensor powers, while an explicit channel attains the bound. This also gives an explicit optimal purification. The same argument extends to finite products of normalized symmetric Werner states with different local dimensions.

The behavior near the symmetric endpoint provides a complementary perspective. The endpoint is also stable in a precise sense: adding an antisymmetric component of weight $\varepsilon$ changes the optimal purity continuously, with variation of order $\sqrt{\varepsilon}$. The canonical purification therefore remains asymptotically optimal in obtained entropy as the symmetric endpoint is approached. This shows that the exact endpoint solution is not isolated at the level of the optimal value, while at the same time it can reveal a nonlinear sensitivity that is invisible from the endpoint formula alone. Some questions are natural to ask as the next step, such that whether the exact optimization can be extended to interior Werner parameters, whether additivity persists away from the symmetric endpoint, or whether optimal purifications can be classified or shown to be unique under additional assumptions. It would also be interesting to understand whether similar constructions apply to other highly symmetric state families or to other R\'enyi orders.

\bigskip
\noindent \textbf{Acknowledgments.}
The author thanks Graeme Smith and John Watrous for invaluable comments, also acknowledges support from the University of Waterloo, Natural Sciences and Engineering Research Council of Canada, and the Government of Canada through the Department of Innovation, Science and Economic Development and by the Province of Ontario through the Ministry of Colleges and Universities at Perimeter Institute.
%

%


\bibliographystyle{unsrt}
\bibliography{sampleQI}

\end{document}